\documentclass[12pt]{iopart}

\usepackage[utf8]{inputenc}
\usepackage{iopams}  
\usepackage{graphicx}  % has \includegraphics, etc.
\expandafter\let\csname equation*\endcsname\relax
\expandafter\let\csname endequation*\endcsname\relax
\usepackage{amsmath}

\usepackage[labelfont={small},subrefformat=parens,caption=false]{subfig}
\usepackage{amssymb,amsbsy,amsfonts,bm}
\usepackage{gensymb}
\usepackage{bbm}
\usepackage{url} 
\usepackage{nicefrac}
\usepackage{hyperref}
\usepackage{xcolor}

\usepackage{dcolumn}
\newcolumntype{d}[1]{D{.}{.}{#1}}
\usepackage{xparse}

\usepackage[normalem]{ulem}
\usepackage{booktabs}  % Pretty tables
\usepackage{multirow}
\usepackage{makecell}
\usepackage{cellspace}
\usepackage{dsfont} 
\definecolor{dgreen}{rgb}{0.0,0.5,0.0}

\begin{document}
	
	\title[Statistical aspects of nuclear mass models]{Statistical aspects of nuclear mass models}

	\author{V. Kejzlar$^1$, L. Neufcourt$^{1,2}$, W. Nazarewicz$^3$, P.-G.~Reinhard$^4$}
	\address{$^{1}$ Department of Statistics and Probability, Michigan State University, East Lansing, Michigan 48824, USA\\
		$^2$ Facility for Rare Isotope Beams, Michigan State University, East Lansing, Michigan 48824, USA\\
		$^3$ Department of Physics and Astronomy and FRIB Laboratory, 
		Michigan State University, East Lansing, Michigan  48824, USA \\
		$^4$ Institut f\"{u}r Theoretische Physik, Universit\"{a}t Erlangen, D-91054 Erlangen, Germany
	}

	\begin{abstract}
We study the information content of nuclear masses from the perspective of global models of nuclear binding energies. To this end, we employ a number of statistical methods and diagnostic tools, including
Bayesian calibration, Bayesian model averaging, chi-square correlation analysis, principal component analysis, and empirical coverage probability. Using a Bayesian framework, we investigate the structure of the 4-parameter Liquid Drop Model by considering discrepant mass domains for calibration. We then use the chi-square correlation framework to analyze the 14-parameter Skyrme energy  density functional calibrated using homogeneous and heterogeneous datasets. We show that quite a dramatic parameter reduction can be achieved in both cases. The advantage of Bayesian model averaging for improving uncertainty quantification is demonstrated. The statistical approaches used are pedagogically described; in this context this work can serve as a guide for future applications.
	\end{abstract}
	
	%%%%%%%%%%%%%%%%%%%%%%%%%%%%%%%%%%%%%%%%%%%%%%%%%%%%%%%%%%%%%%%%%%%%
	%%%%%%%%%%%%%%%%%%%%%%%%%%%%%%%%%%%%%%%%%%%%%%%%%%%%%%%%%%%%%%%%%%%%
	
	\section{Introduction} \label{sec:introduction}

To an increasing extent, theoretical nuclear physics involves statistical inference 
on 
computationally-demanding  theoretical models that often combine heterogeneous datasets. Advanced statistical approaches can enhance the quality of nuclear modeling in many ways \cite{Dob14,ISNET}. First, the statistical tools of uncertainty quantification (UQ) can be used  to estimate theoretical errors on computed observables. Second, they can help to assess the information content of measured observables with respect to theoretical models, assess the
information content of present-day theoretical models with respect to measured observables,
and find the intricate correlations between computed observables -- all
in order to speed-up the cycle of the scientific process. Importantly, they can be used to understand a model’s structure through parameter estimation and model reduction. Finally, statistical tools
can  improve predictive capability and optimize knowledge extraction  by extrapolating beyond the regions reached by experiments to provide meaningful input to applications and planned measurements. 

In this context, Bayesian machine learning \cite{KoH} 
 can address many of these issues in a unified and comprehensive way by combining the current-best theoretical and experimental inputs into a quantified prediction, see Refs. 
\cite{McDonnell2015,Higdon2015,Neufcourt2018,Neufcourt2019,Neufcourt2020,Neufcourt2020a,Sprouse2019} for relevant example of Bayesian studies pertaining to nuclear density functional theory (DFT) and  nuclear masses (see also Refs.~\cite{Utama16,Utama17,Utama18,Niu2018,Rodrguez2019} on Bayesian neural network applications to nuclear masses).

To demonstrate the  opportunities in nuclear theory offered by statistical tools, we carry out in this study 
the analysis of two nuclear mass models informed by measured masses of even-even nuclei. We begin with the semi-empirical mass formula given by the Liquid Drop Model (LDM)  \cite{Weizsacker1935,Bethe36,Myers1966,Kirson2008,Benzaid2020} whose parameters are obtained by a fit
to nuclear masses. Because of its linearity and  simplicity, 
the LDM has become a popular model for various statistical applications \cite{Bertsch2005,Toivanen2008,Utama16,Yuan2016,Bertsch2017,Zhang2017,Cauchois2018,Shelley2019,Pastore2019,Carnini2020}. We study the impact of the fitting domain on the parameter estimation of the LDM, which is carried out by means of both chi-square and Bayesian frameworks. To learn about the number of effective parameters of the LDM, we perform a principal component analysis. By combining LDM parametrizations optimized to light and heavy nuclei, 
we demonstrate the virtues of Bayesian model averaging.

In the second part of our paper, we check the LDM robustness by investigating the structure of the more realistic Skyrme energy density functional. By means of the chi-square correlation technique, we study different Skyrme parametrizations obtained by parameter optimization using homogeneous and  heterogeneous datasets. Finally, we perform a principal-component analysis of the Skyrme functional to learn about the number of its effective degrees of freedom. 

In the context of the following discussion, it is useful to 
 clarify the  notion of a ``model''. In this work, by model we understand the combination of a raw theoretical model (i.e., mathematical/theoretical framework), the calibration dataset used for its parameter determination,
 and a statistical model that describes the error structure.

%%%%%%%%%%%%%%%%%%%%%%%%%%%%%%%%%%%%%%%%%%%	
\section{Liquid Drop Model in different nuclear domains}
%%%%%%%%%%%%%%%%%%%%%%%%%%%%%%%%%%%%%%%%%%%	

The semi-empirical mass formula of the LDM parametrizes the binding energy of the nucleus
$(Z, N)$ as:
	\begin{equation}\label{eqn:LDM}
	E_{\rm LDM}(N,Z) = a_{\rm vol}A - a_{\rm surf}A^{2/3} - a_{\rm sym} \frac{(N-Z)^2}{A} - a_{\rm C} \frac{Z(Z-1)}{A^{1/3}},
	\end{equation}
where $A=Z+N$ is the mass number and the successive terms
represent the volume, surface, symmetry, 
and Coulomb energy, respectively. 
The expression (\ref{eqn:LDM}) can be viewed
in terms of the binding-energy-per-nucleon expansion in 
terms of powers of $A^{-1/3}$ (proportional to inverse radii)  
and the squared neutron excess
(related to the neutron-to-proton asymmetry $(N-Z)/A$). 
This kind of  expansion, often referred to as leptodermous expansion \cite{Grammaticos1982,Brack1985,Reinhard2006}, should be viewed in the asymptotic sense \cite{Brack1997}. 

At this point, it is worth noting that the quantal shell energy responsible for
oscillations of the nuclear binding energy with particle numbers
 scales with mass number  as $A^{1/3}$, i.e., it scales linearly with the nuclear radius.
 The shell energy is ignored in the macroscopic LDM; it is accounted for  in the  microscopic Skyrme DFT approach  -- discussed later in the paper --  which  is rooted in the concept of single-particle orbitals forming the nucleonic shell structure. In general, the performance of the LDM gets better in heavy nuclei as compared to the light systems,  which are greatly driven by surface effects \cite{Reinhard2006,Yuan2016,Benzaid2020}.
  
\begin{figure}[htb!]
		\center
		\includegraphics[width=0.8\linewidth]{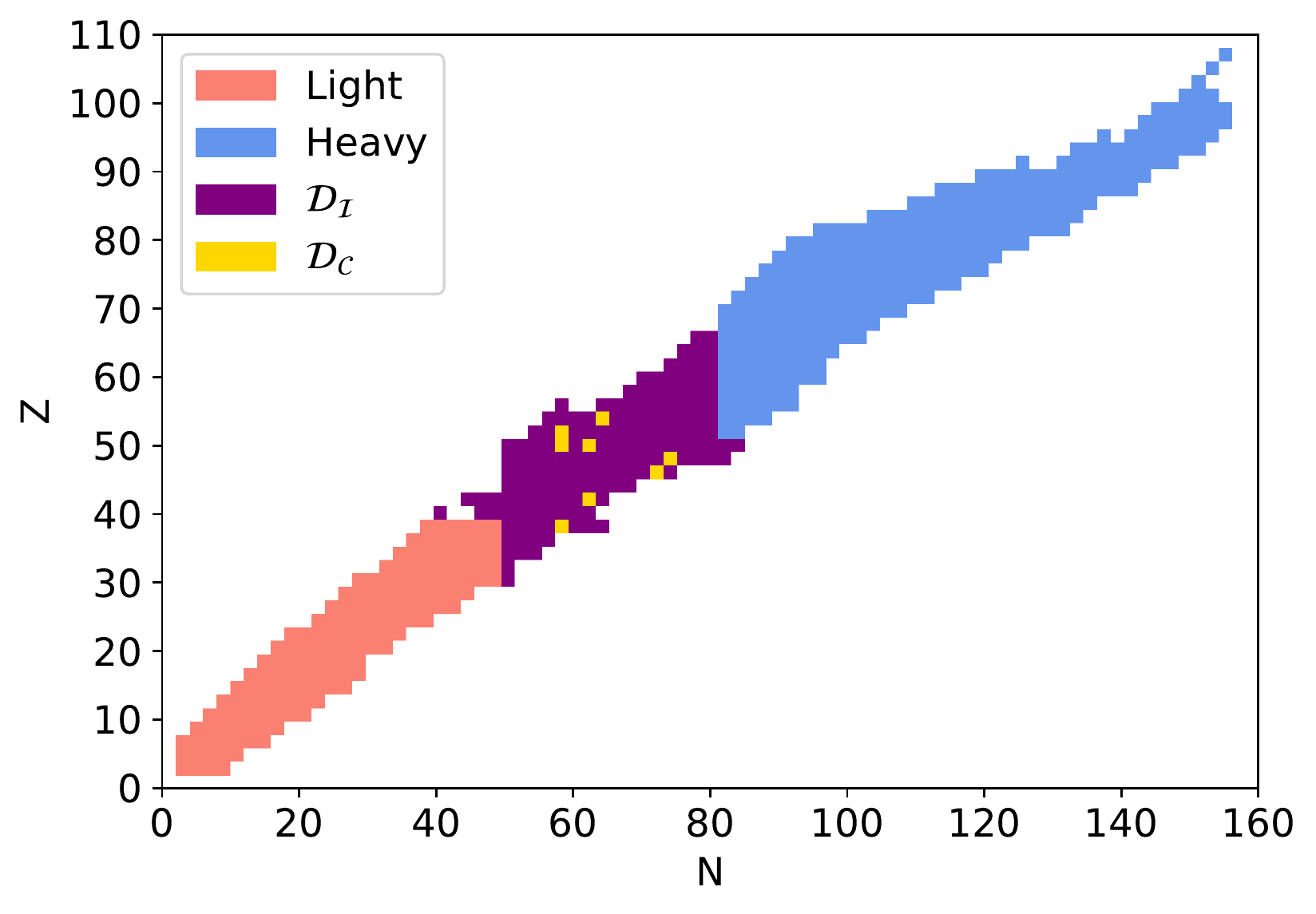}
		\caption{\label{LDM_variants}
			Even-even nuclei from AME2003 divided into the domains of light ($Z<40$, $N<50$), heavy ($Z>50$, $N>80$), and intermediate nuclei (remaining 155 nuclei) denoted $\mathcal{D}_\mathcal{I}$ containing the
subset $\mathcal{D}_\mathcal{C}$ for special counterchecks.
		}
	\end{figure}

To study the impact of the fitting domain on parameter estimation, prediction accuracy, and UQ fidelity of nuclear mass models, 
we shall consider the experimental binding energies of 595 even-even nuclei of AME2003 divided into 3 domains according to Fig. \ref{LDM_variants}. Namely, we define the domain of \textit{light} nuclei with $Z < 40$ and $N< 50$, \textit{heavy} nuclei with $Z > 50$ and $N > 80$, and the \textit{intermediate} domain $\mathcal{D}_\mathcal{I}$ consisting of  the remaining even-even nuclei. To keep some of our results within computable ranges
we will also consider 8 randomly selected nuclei in the \emph{central} subset 
 of the intermediate domain which we will denote $\mathcal{D}_\mathcal{C}$.
By dividing nuclear domains according to $A$, 
we are trying to 
simulate the current theoretical strategy in modeling atomic nuclei: light nuclei are often described by different classes of models ($A$-body models) than heavy nuclei (configuration interaction, DFT), with the intermediate domain being the testing ground for all approaches \cite{Nazarewicz2016}. Here we use, for testing, the same LDM
  expression in all domains. The models are distinguished merely by
  the fitting datasets.
Let us emphasize that  in realistic nuclear physics applications, light and heavy nuclei are usually treated by means of different raw theoretical frameworks \emph{and} different calibration datasets.  It is only for the sake of this  pedagogical manuscript, to see clearly the statistical outcomes,  that we decided to distinguish between  different LDM variants by considering different  fitting domains only.

%%%%%%	
%%%%%%

In terms of these separated data domains,  we consider four LDM variants fitted on specific regions of the nuclear landscape:
\begin{description}
		\item[(i)] LDM(A) -- LDM fitted on all 595 even-even nuclei.
		\item[(ii)]  LDM(L) -- LDM restricted to the {light} domain (153 nuclei).
		\item[(iii)]  LDM(H) -- LDM restricted to the {heavy} domain (287 nuclei).
		\item[(iv)]  LDM(L + H) -- LDM fitted on the both {light} and {heavy} domain (440 nuclei).
\end{description}
We emphasize that the {intermediate} domain $\mathcal{D_I}$
(and \emph{a fortiori} $\mathcal{D_C}$)
is not used for training in variants (ii)-(iv),
but kept aside as an independent testing domain
where the different LDM variants compete.
Thus we use the binding energies in the {intermediate} domain to evaluate
the predictions and error bounds of these variants and their Bayesian averages.
In short, this setup is designed to produce a scenario where 
two models, which have been optimized on their 
respective domains, compete to explain the data on a  
third disconnected domain. 

 Note that, rigorously, there are two possible violations of independence. First, some of the experimental values from the test set may come from the same measurement as other values in the training set. Second, in the absence of a finer treatment of systematic errors, 
 where autocorrelations have been shown to range over 4-9 particle numbers, these are also contained in the $\sigma$ $\epsilon_i$ term of the model.
Thus, we can consider that the domains $D_C$ and $D_I$ are, respectively, large enough and far enough from the training sets to guarantee approximate independence.

We would like to point out in passing that fitting binding energy per nucleon corresponds to a radically different model 
from a statistical perspective as it relies 
on a different assumption on the structure of the errors. A
simple analysis shows that the scaling of the LDM residuals 
is relatively more uniform with $A$
when considering binding energy 
than when considering binding energy per nucleon, 
confirming that fitting binding energies is the correct approach.

    %%%%%%%%%%%%%%%%%%%%%%%%%%%%%%%%%%%%%%%%%%%%%%%%%%%
	\section{Liquid Drop Model: parameter estimation}
    %%%%%%%%%%%%%%%%%%%%%%%%%%%%%%%%%%%%%%%%%%%%%%%%%%%
    
In this section we compare the results of traditional chi-square fit and Bayesian calibration. We also explore the possibility of reducing the LDM parameter space  via a principal component analysis. 
	
Our statistical model for binding energies $y_i$ can be written as:
	\begin{equation}\label{eqn:stat_model}
	y_i = f(x_i, \theta) + \sigma\epsilon_i,
	\end{equation}
where the function $f(x, \theta)$ represents the LDM  prediction \eqref{eqn:LDM} with a given parameter vector $\theta = (a_{\rm vol},a_{\rm surf}, a_{\rm sym}, a_{\rm C})$ for a nucleus indexed by $x = (Z,N)$. The  errors are modeled as independent standard normal random variable $\epsilon_i$  with mean zero and unit variance, scaled by an adopter error $\sigma$
that reflects the model's incapability to  follow the data (which, in the context of nuclear mass models,  is usually much greater than the experimental error).

While the function $f$ is nonlinear in $x$,
it is linear in the parameter vector $\theta$,
thus this model falls conveniently in the family of
the generalized linear models (GLM) \cite{khuri2009linear} and can be treated 
by means of a standard linear regression.
	
	%%%%%%%%%%%%%%%%%%%%%%%%%%%%%%%%%%%%%%%%%%%%%%%%%%%
	\subsection{Chi-square analysis}
    %%%%%%%%%%%%%%%%%%%%%%%%%%%%%%%%%%%%%%%%%%%%%%%%%%%
	
Given the datapoints $y_i$ for $i=1...N$, we define the estimate $\hat\theta$ of the parameter vector $\theta$ as the minimizer of the penalty function
	\begin{equation}\label{eqn:chisq}
	\chi^2(\theta)
	=
	\sum_{i=1}^N\left(y_i - f(x_i, \theta) \right)^2.
	\end{equation}
This optimization problem has a closed form solution for  functions $f$ linear in $\theta$ (this is true for the LDM)
\cite{khuri2009linear} given by the maximum likelihood estimator
	\begin{equation}\label{eqn:OLS}
	\hat\theta = (\bm{J}^T\bm{J})^{-1}\bm{J}^TY,
	\end{equation}
where $Y$ is a column vector of datapoints $y_i$ and $\bm{J}$ is the Jacobian:
	\begin{equation}\label{eqn:jacobian}
	J_{i\alpha} = \frac{\partial f(x_i, \theta)}{\partial \theta_\alpha}.
	\end{equation}
The assumption of Gaussian  error in \eqref{eqn:stat_model} implies for the true value 
of $\theta$ the probability distribution
	\begin{equation}
	\mathcal{P}(\hat\theta) \propto \exp{\left(-\frac{1}{2\sigma^2}(\hat\theta - \theta)^T\bm{H}(\hat\theta - \theta)\right)},
	\label{eq:param_dist}
	\end{equation}
where $\bm{H}$ is the Hessian with elements defined as 
	\begin{equation}\label{eqn:hess_elem}
	H_{\alpha \beta} = \frac{\partial^2\chi^2(\theta)}{\partial\theta_\alpha\partial\theta_{\beta}}\Bigr|_{\substack{\theta=\hat\theta}}\,.
	\end{equation}
The same expression holds  for a general function $f$ 
beyond the GLM framework 
to the extent that it can be reasonably approximated by its first-order Taylor expansion around $\hat\theta$.

For the subsequent  principal component analysis, we  introduce dimensionless model parameters $\widetilde{\theta}_\alpha$ as
	\begin{equation}\label{eqn:dimensionless}
	\widetilde{\theta}_\alpha  := \frac{\theta_\alpha}{\delta\theta_\alpha},
	\end{equation}
where we use the uncorrelated variance of a parameter, $\delta \theta_\alpha = 1/{(\sigma\sqrt{H_{\alpha \alpha}})}$ \cite{Bevington}, to define a natural
scale that changes the setup to what we call conditioned Hessian distinguished by a tilde: 
	\begin{equation}
	\widetilde{H}_{\alpha \beta}= \frac{H_{\alpha\beta}}{\sqrt{H_{\alpha \alpha}}\sqrt{H_{\beta \beta}}}.
	\end{equation}
If the root-mean-square (rms) error is known \textit{a priori}, the scaling parameter $\sigma$ can be fixed. Otherwise, it can be estimated from the data as 
    \begin{equation}\label{eqn:error}
     {\sigma} = \sqrt{\frac{\sum_{i=1}^N\left(y_i - f(x_i, \hat\theta) \right)^2}{N-p}}, 
    \end{equation}
 where $p$ is the number of parameters in the model ($p = 4$ in the case of LDM).
	
Given the distribution (\ref{eq:param_dist}), the correlated variance of the parameter estimate is expressed through the covariance matrix $\bm{C}=\sigma^2 \bm{H}^{-1}$ \cite{Bevington}. The same holds for the
dimensionless parameters associated  with the conditioned covariance matrix
$\widetilde{\bm{C}}=\sigma^2 \widetilde{\bm{H}}^{-1}$. The diagonal elements of $\widetilde{\bm{C}}$
represent the correlated error on the model parameters:
	\begin{equation}
	\Delta\tilde{\theta}_\alpha
	=
	\sqrt{\tilde{C}_{\alpha\alpha}}
	\;.
	\end{equation}

The normalized covariance matrix
	\begin{equation}\label{Pearson}
	c_{\alpha\alpha'}
	=
	\frac{C_{\alpha\alpha'}}{\sqrt{C_{\alpha\alpha}C_{\alpha'\alpha'}}}
	\end{equation}
quantifies the degree of alignment (correlation) between $\alpha$ and $\alpha'$. The quantity  $|c_{\alpha\alpha'}|^2$ is called  the coefficient-of-determination (CoD); it  is another way of representing the correlation between the two model parameters \cite{Rei13e,Dob14}. The matrix of CoDs is  positive semidefinite. A value  $|c_{\alpha\alpha'}|^2=1$ stands for the complete correlation and  $|c_{\alpha\alpha'}|^2=0$ for the full independence of parameters.
	
	\begin{table}[htb!]\centering
		\caption{Parameter estimates $\hat\theta$ for LDM variants (i) - (iv), with corresponding correlated errors $\Delta\theta$ (all in MeV). 
		The results are based on the unconditioned covariance matrix $\bm{C}$.}
		\vspace{1mm}
		\renewcommand{\arraystretch}{1.15}
		\begin{tabular}{crrrrrrrrr}
			\hline 
			\hline 
			& \multicolumn{2}{c}{LDM(A)} & \multicolumn{2}{c}{LDM(L)} & \multicolumn{2}{c}{LDM(H)} & \multicolumn{2}{c}{LDM(L+H)} \\
			\multicolumn{1}{l}{~~$\theta$} & \multicolumn{1}{l}{~~~$\hat\theta$} & \multicolumn{1}{l}{~~$\Delta\theta$} & \multicolumn{1}{l}{~~~$\hat\theta$} & \multicolumn{1}{l}{~~$\Delta\theta$} & \multicolumn{1}{l}{~~~$\hat\theta$} & \multicolumn{1}{l}{~~$\Delta\theta$} & \multicolumn{1}{l}{~~~$\hat\theta$} & \multicolumn{1}{l}{~~$\Delta\theta$} \\
			\hline
			$a_{\rm vol}$ & 15.162 & 0.051 & 14.050 & 0.097 & 15.221 & 0.176 & 15.162 & 0.057 \\
			$a_{\rm surf}$ & 15.960 & 0.160 & 13.877 & 0.230 & 15.873 & 0.624 & 16.002 & 0.174 \\
			$a_{\rm sym}$ & 21.995 & 0.131 & 17.054 & 0.347 & 22.502 & 0.390 & 22.037 & 0.151 \\
			$a_{col}$ & 0.680 & 0.004 & 0.534 & 0.013 & 0.690 & 0.010 & 0.679 & 0.004 \\
			$\sigma$ & 3.698 &  & 2.936 &  & 2.690 &  & 3.841 & \\
			\hline
			\hline
		\end{tabular}
		\label{tab:LDM_freq_est}
	\end{table}
The LDM parameter estimates corresponding to the minimization of the chi-square penalty on the varying domains of the data are displayed in Table~\ref{tab:LDM_freq_est}. There are significant differences between the parameter values of the light and heavy variants, in particular the ones associated with lowest order corrections; volume energy is higher when fitted to the heavy nuclei than to the light ones. 
This is not surprising as the compensation between volume and surface terms is significant for the light nuclei.
As could be expected, the  parameters obtained in the two combined variants fall in between. Taking LDM(A) as a reference, these parameter estimates fall within one-sigma error bars of LDM(H) and LDM(L+H). They are however inconsistent with the LDM variant fitted to the set of {lighter} nuclei -- falling outside of its five-sigma error bars.
Comparing the $\Delta\theta$ values 
for LDM(A) and LDM(L+H) with those of LDM(L) and LDM(H),
one can also notice that the correlated errors of the parameters are significantly reduced with the size of the  dataset.
	
In addition to the model parameter $\theta$ itself, the error scaling parameter $\sigma$ also varies with the  domain, from $2.69$ MeV for LDM(H) to $3.84$ for LDM(L+H). This indicates that the residuals of the LDM are not uniformly distributed with respect to the values of $Z$ and $N$. This is to be expected: the  leptodermous expansion becomes more accurate for heavy nuclei  \cite{Reinhard2006}, which are dominated by the volume effects.

	\begin{figure}[htb]
		\center
		\includegraphics[width=0.8\linewidth]{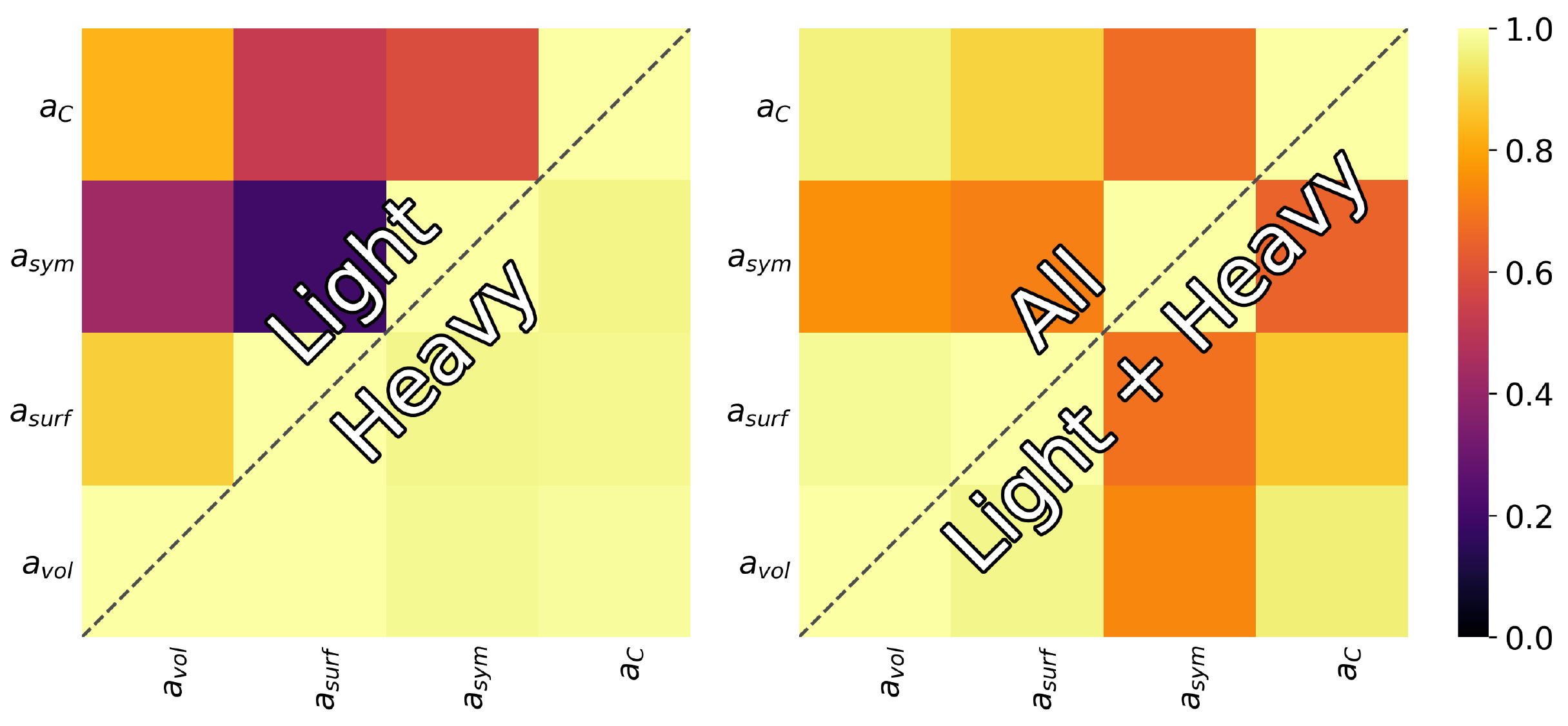}
		\caption{\label{fig:LDMcorr}
			Matrices of CoD for LDM variants (i)-(iv) optimized to the masses of even-even nuclei from AME2003. 
		}
	\end{figure}
%%%%	
	Another  insight into the structure of the LDM can be obtained from the correlations between the model parameters shown in Fig.~\ref{fig:LDMcorr} in terms of CoDs. Here, particularly instructive is the comparison between LDM(L) and LDM(H). In the case of the heavy nuclei, all LDM parameters are extraordinarily well correlated. For the light nuclei, the correlation between symmetry and Coulomb terms is small as the ranges of neutron excess and atomic numbers are limited, and their correlations with the volume and surface terms deteriorate. When the analysis is performed on the large datasets of LDM(A) and LDM(L+H), all parameters are highly correlated as recently noticed in Refs.~\cite{Shelley2019,Pastore2019,Cauchois2018}. The lesson learned from Fig.~\ref{fig:LDMcorr} is that the choice of a fit-dataset does impact inter-parameter correlations.  This can have consequences on the generality of the model and potentially reduce
 its predictive performance \cite{kutner2005applied}.
As discussed later, the pattern of parameter correlations  can be strongly
influenced by the use of heterogeneous datasets in which fit-observables can be grouped into different classes (masses, radii, etc.).

\begin{table}[htb!]\centering
		\caption{
			RMSDs (in MeV) of the predictions 
			from the 4 LDM variants 
			as well as the values from the Bayesian model averaging described in  Sec.~\ref{sec:BMA},
			calculated on the held-out data in the intermediate domain of even-even nuclei from AME2003,
			using the chi-square and the Bayesian calibrations.
		}		\label{tab:rms}
		\vspace{1mm}
		\renewcommand{\arraystretch}{1.15}
		\begin{tabular}{l|lcccc}
			\hline
			\hline
			&  & LDM(A) & LDM(L) & LDM(H) & LDM(L+H)\\
			\hline
\multirow{4}{*}{$\mathcal{D}_\mathcal{I}$} & chi-square & 3.205 & 8.170 & 3.817 & 3.351 \\
& Bayes & 3.206 & 8.176 & 3.811 & 3.351 \\
\cline{2-6}
& BMA(L,H) & \multicolumn{4}{c}{3.810}\\
& BMA(L,H,L+H) & \multicolumn{4}{c}{3.223}\\
\hline
\hline
\multirow{4}{*}{$\mathcal{D}_\mathcal{C}$} & chi-square & 1.930 & 6.817 & 3.307 & 1.879 \\
& Bayes & 1.930 & 6.825 & 3.292 & 1.881 \\
\cline{2-6}
& BMA(L,H) & \multicolumn{4}{c}{3.300}\\
& BMA(L,H,L+H) & \multicolumn{4}{c}{1.926}\\
			\hline
			\hline
		\end{tabular}
	\end{table}
	
We assess the predictive performance of the chi-square fit of the four LDM variants by using the root-mean-square deviation (RMSD):
	\begin{equation}
	{\rm RMSD} = \sqrt{\frac{\sum_{i=1}^n\left(y^*_i - f(x^*_i, \hat{\theta}) \right)^2}{n}},
	\end{equation}
calculated on experimental binding energies $y^*_i$ in the held-out data in the intermediate domain with $40\le Z \le 50$ and $50\le N \le 80$, used as an independent testing dataset. Results are shown in the line of Table \ref{tab:rms} denoted chi-square. As expected, the LDM(A) variant fitted to all the even-even nuclei  performs the best with 
RMSD around 3.2\,MeV. The  performance of
LDM(L), having the largest RMSD of 8.17\,MeV, is poor. As compared to LDM(A), there is only a small loss in the predictive power for  LDM(H) and LDM(L+H), which both  compete meaningfully on the intermediate domain.
	
	\subsection{Bayesian calibration}\label{BAnal}
	
The Bayesian approach consists here of looking at the (full) posterior distribution of $(\theta, \sigma)$ given by Bayes's rule:
	\begin{equation}
	p(\theta, \sigma | y) \propto p(y|\theta, \sigma) \pi(\theta, \sigma),
	\end{equation}
where $p(y|\theta, \sigma)$ is the model likelihood given by \eqref{eqn:stat_model} and $\pi(\theta, \sigma)$ is the prior distribution on the parameters $\theta$ and the error $\sigma$.

We shall use in this study 
\emph{weakly informative} priors, 
i.e., arbitrary distributions where hyperparameters are chosen
to ensure that the prior distribution spans a much wider domain than the resulting posterior. 
The theoretical bias introduced by these priors can be removed using a non-informative prior, scaled according to the variations of the data  \cite{gelman2013bayesian}. A recent study shows how this can be done  without using usual infinite dataset approximations \cite{Mattingly2018}. For practical purposes, weakly informative priors are non-informative in essence.
Gaussian priors are typical choices for non-constrained parameters;
as for (non-negative) noise scale parameters 
common defaults include half-normal, half-Cauchy, Inverse Gamma and Gamma distributions -- a family including chi-square distributions
\cite{lunn2012bugs, gelman2013bayesian}.

Consequently, for the LDM parameters 
$a_{\rm vol}, a_{\rm surf}$ and $a_{\rm sym}$ we use
independent normal prior distributions ${\cal N}(0,100)$
with mean 0 and standard deviation 100, while for $a_{\rm C}$ we take ${\cal N}(0,2)$.
For $\sigma$ we assume a gamma prior distribution $\Gamma(5,0.5)$ with shape parameter 5 and rate parameter $0.5$ (thus mean 10 and variance 20).

Similarly to the chi-square fit, the scale parameter $\sigma$ can be also fixed to an \emph{a priori} value, in which case the posterior distribution of interest is $p(\theta|y)$. In general, samples can be conveniently obtained from an ergodic Markov chain produced by the Metropolis-Hastings algorithm, an extension of the Gibbs sampler \cite{gelman2013bayesian,Gilks}.
	
	\begin{figure}[htb!]
		\center
		\includegraphics[width=0.7\linewidth]{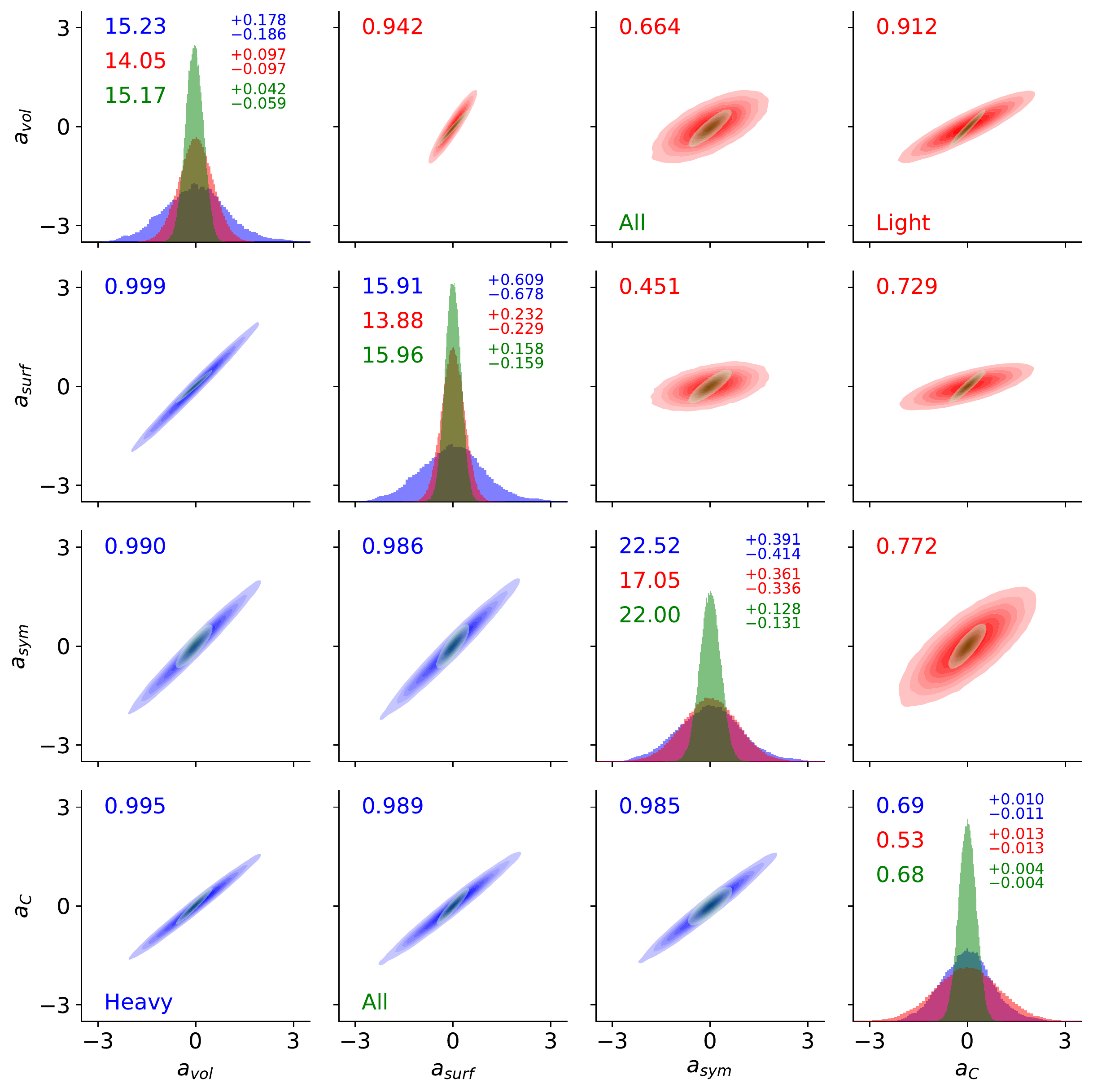}
		\caption{\label{fig:LDM_bayes}
			Posterior distributions of the model parameters for LDM variants LDM(A), LDM(L), and LDM(H) obtained from MCMC samples. Upper triangle corresponds to LDM(L), lower triangle to LDM(H), and the LDM(A) fit is overlaid on both off-diagonals. Before being plotted, posterior samples for each of the parameters were centered at zero and scaled by the largest sample deviation for a given parameter among all the LDM variants. Posterior means and $68\%$ HPD credible intervals are indicated by numbers, as are correlation coefficients \eqref{Pearson} for all parameter pairs.
}
	\end{figure}	
Informed predictions for the binding energies $y^*$ in the intermediate domain are given by the posterior predictive distribution $p(y^*|y)$. This can be produced from the posterior parameter distributions  by integrating the conditional density of $y^*$, given $(\theta, \sigma)$ and the training binding energies $y$, against the posterior density $p(\theta, \sigma | y)$: 
\begin{equation}\label{eqn:predictive_density} 
p(y^*|y) = \int p(y^*| y, \theta, \sigma) p(\theta, \sigma | y)d\theta d\sigma. 
\end{equation}
The conditional density $p(y^*| y, \theta, \sigma)$ is again given directly by the statistical model \eqref{eqn:stat_model}. The assumption of independent error $\epsilon_i$ yields $p(y^*| y, \theta, \sigma) = p(y^*| \theta, \sigma)$. In other words, the value of $y^*$ is conditionally independent of $y$ given the statistical model parameters $\theta$ and $\sigma$. It is also worth noting that the posterior predictive density is rarely computed directly from Eq.~\eqref{eqn:predictive_density}. Instead, if samples $(\theta^{(1)}, \sigma^{(1)}),\dots,(\theta^{(M)}, \sigma^{(M)})$ are produced from the posterior density $p(\theta, \sigma | y)$ via a Monte Carlo Markov chain (MCMC), the corresponding samples $y^{*(1)}, \dots, y^{*(M)}$  follow the posterior density $y^{*(i)} \sim p(y^*| \theta^{(i)}, \sigma^{(i)})$, $i = 1, \dots, M$. The posterior predictive density is then approximated using the empirical density of samples $y^{*(1)}, \dots, y^{*(M)}$.

Together with posterior mean predictions,
we can extract from the posterior samples
Highest Posterior Density (HPD)  credible intervals -- the Bayesian counterpart to frequentist confidence intervals.
Given a credibility level $\alpha$, the $\alpha$-HPD of a scalar quantity consists of the minimum width interval containing an $\alpha$ fraction of its MCMC posterior samples.
We will consider in particular the $68\%$ HPD credible intervals which mimic the frequentist ``one-$\sigma$'' error bars.

\begin{figure}[htb!]
	\center
	\includegraphics[width=0.7\linewidth]{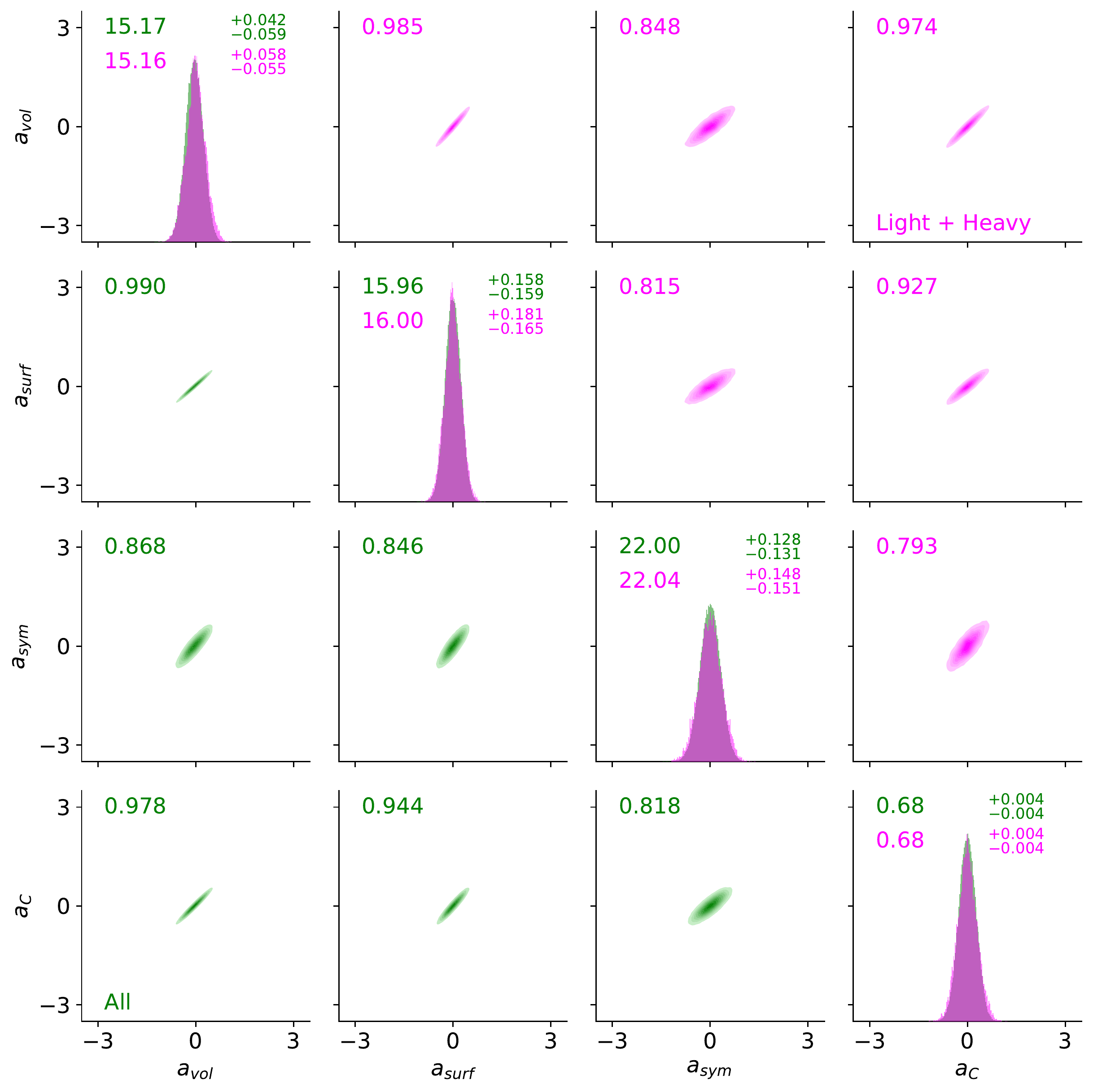}
	\caption{
	    \label{fig:LDM_bayes_A_LH}
        Similar as in Fig.~\ref{fig:LDM_bayes} but for the LDM(A) and LDM(L+H) variants. The scaling of the posterior samples is consistent with 
        Fig.~\ref{fig:LDM_bayes}.
    }
\end{figure}	
	
Figures~\ref{fig:LDM_bayes} and \ref{fig:LDM_bayes_A_LH} show the bivariate posterior distributions of the LDM model based on $2\times 10^5$ MCMC samples obtained using the modern No-U-Turn MCMC sampler \cite{NUTS}. Due to a nearly-Gaussian behavior of the posterior distributions,  the posterior means are very close to the $\hat\theta$ values in Table~\ref{tab:LDM_freq_est} obtained in the chi-square analysis. In fact, they all coincide within the one-sigma error bar. This shows practical equivalence of the linear regression technique and Bayesian analysis when it comes to the LDM parameter estimation.

As discussed in Fig.~\ref{fig:LDMcorr} in the context of chi-square analysis,  there is a general positive correlation between all the parameters for all models. It is particularly strong for LDM(H) ($>98\%$) with lesser, yet still strong, correlation for LDM(A) and LDM(L+H)  ($>80\%$). The lowest correlations are between $a_{\rm sym}$ and the volume and surface coefficients in LDM(L), see Fig.~\ref{fig:LDM_bayes}. 
	
\begin{figure}[htb!]
		\center
		\includegraphics[width=0.6\linewidth]{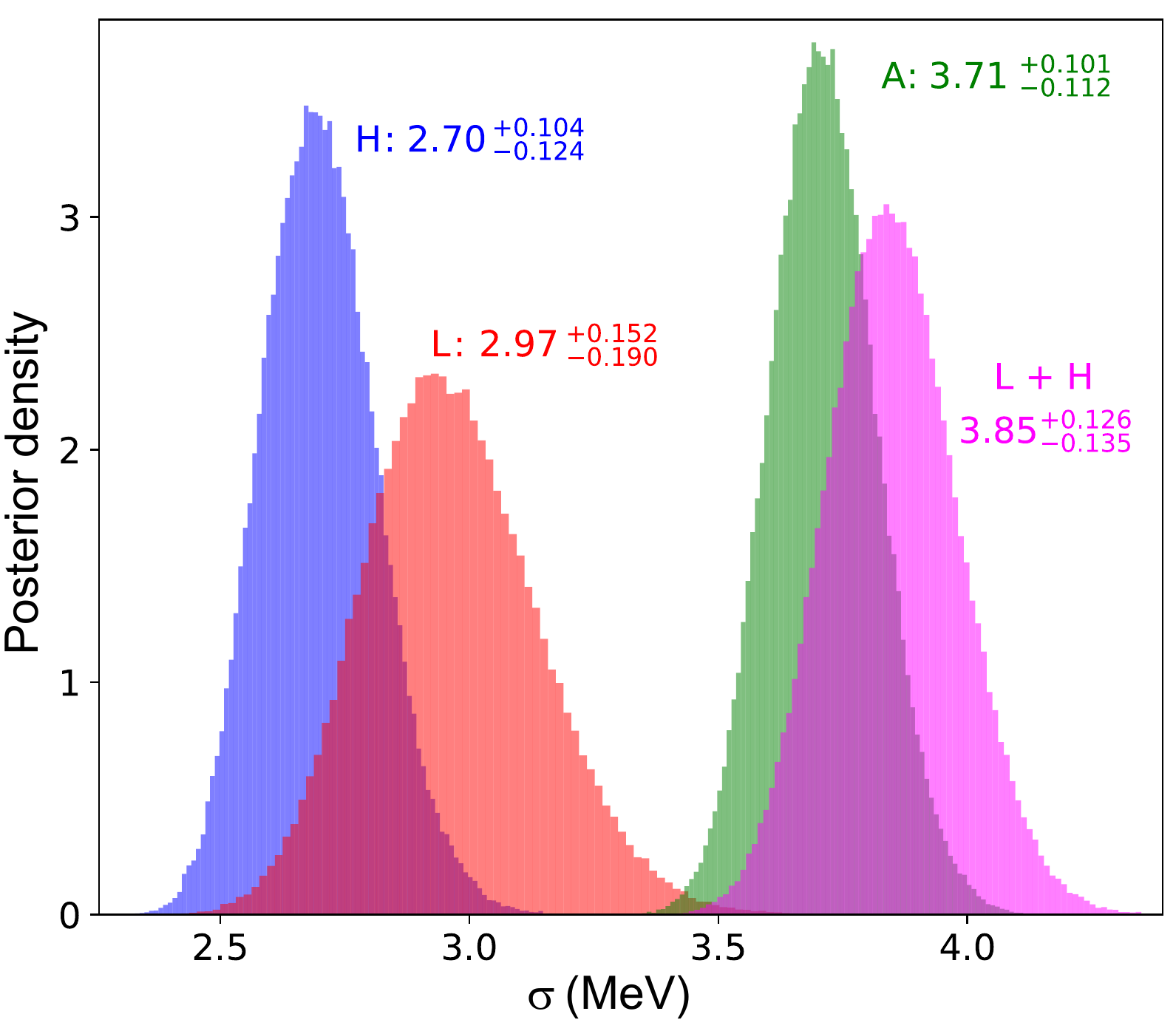}
		\caption{\label{fig:LDM_bayes_sigma}
			Posterior distribution of scale parameter $\sigma$ for LDM variants (i) -(iv).  Posterior means and $68\%$ HPD credible intervals are indicated by numbers.
		}
	\end{figure}

We also show in Fig. \ref{fig:LDM_bayes_sigma} the posterior distribution of the scale parameter $\sigma$ for the four LDM variants. The posterior means are relatively close ($2.70 -3.85$\,MeV).  The models fitted on a large dataset (A, L+H) produce higher values of $\sigma$ as they try to accommodate masses of both the light and heavy nuclei. Indeed, one can interpret this by considering that posterior samples conditioned on the combined domain 
incorporate part of the uncertainty tied to the model.
%%%%
	
The RMSDs obtained from the Bayesian calibration 
(corresponding to the predictions based on the posterior mean 
of the parameters)
are  displayed 
in Table \ref{tab:rms}. We see that these values are practically identical to those obtained in the chi-square analysis.

	\subsection{Principal component analysis}
	
The idea beyond principal component  analysis  is to transform a set of variables (here: parameters) into a set 
of linearly uncorrelated components, 
with  the first (principal) component accounting for as much of the variability as possible \cite{Pearson1901,ELSII,Jolliffe2002}. In practice, this is achieved by carrying out the  singular value decomposition (SVD) of the
conditioned Hessian $\tilde{\bm{H}}$. In the examples considered here, the SVD can be reduced to a diagonalization:
	\begin{equation}\label{princ}
	\sum_{k'}\tilde{H}_{kk'}V_{k'n}=\tilde{h}_nV_{kn}
	\;.
	\end{equation}
The eigenvalues $\tilde{h}_n$ contain the information about redundancy or effective degrees of freedom. The eigenvectors $V_{kn}$ (principal components) contain the parameter correlations, but are often too involved to help an interpretation. The eigenvalues $\tilde{h}_n$ quantify the relevance of an effective parameter $\tilde{\gamma}_n=\sum_k\tilde{\theta}_kV_{kn}$ associated with the principal component $n$. Large $\tilde{h}_n$ means that this principal component has a large impact on the penalty function $\chi^2$ while very small eigenvalues indicate irrelevant parameters having little consequences for the parameter estimation (the penalty function is soft along this direction). One way to weigh the importance of an eigenvalue is the partial-sum criterion \cite{Jolliffe2002}. To this end, one sorts $\tilde{h}_n$ in decreasing order and  requests that the cumulative value
	\begin{equation}
	S_m
	=
	\frac{\sum_{n=1}^m\tilde{h}_n}{\sum_{n=1}^{p}\tilde{h}_n}
	\label{eq:exhaust}
	\end{equation}
	lies above a certain threshold $S_\mathrm{limit}$. A typical setting for that is
          $S_\mathrm{limit}=0.99$, i.e.,  the partial sum is
          exhausted by 99\%.  We note that since the diagonal matrix elements of  the  conditioned Hessian are all equal to one, and 
	$\det(\tilde{\bm{H}})\ne 0$ for practical cases, the sum in the denominator 
	of Eq.~(\ref{eq:exhaust}) is
	$\sum_{n=1}^{p}\tilde{h}_n=\tr(\tilde{\bm{H}})=p$.

For the sake of the following discussion, 	it is useful to consider two trivial limiting cases:
	\begin{description}
		\item[C1]
		No correlation between model parameters. (This corresponds to a perfect choice of a model's degrees of freedom and a maximally peaked likelihood with credibility intervals of minimal size.): $\tilde{\bm{H}}=\mathbb{I}$. In this case, $\tilde{\bm{H}}$ has $p$ eigenvalues equal to 1 and the principal components are in the direction of model parameters;
		\item[C2]
		Perfect correlation between model parameters: $\tilde{\bm{H}}_{i,j}=1~~\forall i,j$. Here,  $\tilde{\bm{H}}$ has $p-1$ eigenvalues equal to 0 and one eigenvalue $\tilde{h}_1=p$ with the eigenvector 
		\begin{equation}\label{V1}
		\bm{V}_1=\frac{1}{\sqrt{p}}[1,1,\cdots, 1].
		\end{equation} 
	\end{description}
Case C1 suggests a lower limit
  for $S_\mathrm{limit}$. Namely, it must be lager than $(p-1)/p$ to cope
  properly with the no-correlation case.

%%%%%
	\begin{figure}[htb!]
		\centerline{\includegraphics[width=1\linewidth]{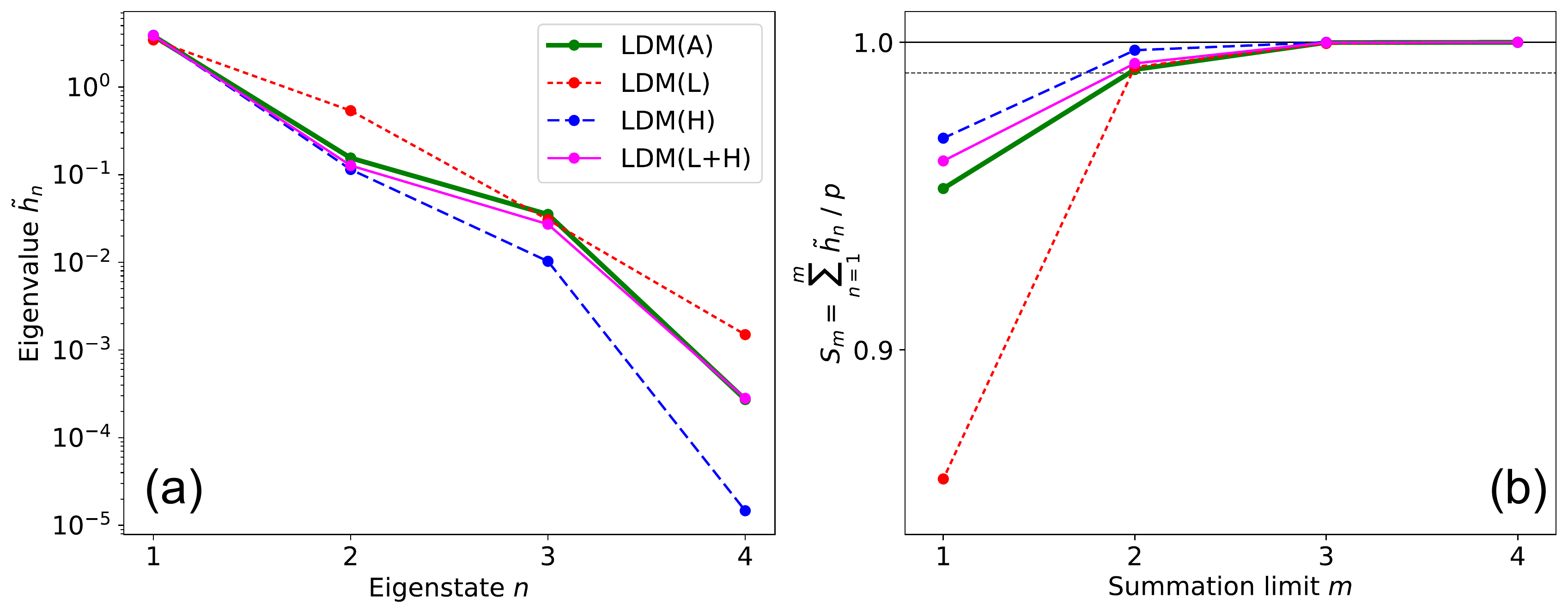}}
		\caption{\label{fig:LDM_svd}
			(a) Eigenvalues $\tilde{h}_n$ of the conditioned Hessian matrix 
			and (b) cumulative percentage (\ref{eq:exhaust})
			for LDM variants (i) -(iv). A dotted horizontal line in (b) indicates the threshold $S_\mathrm{limit}=0.99$.
		}
	\end{figure}	

\begin{figure}[htb!]
		\centerline{\includegraphics[width=0.5\linewidth]{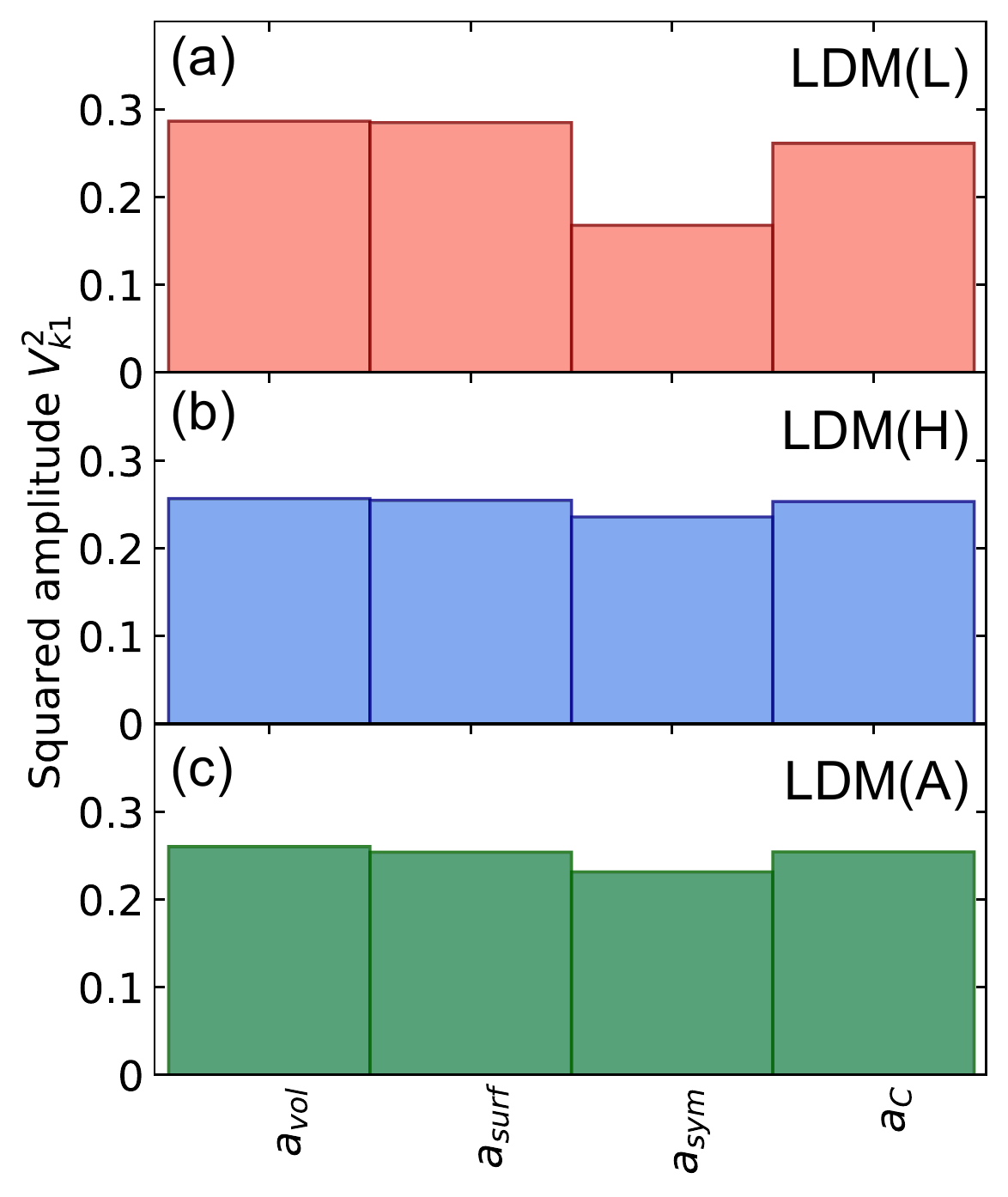}}
		\caption{\label{fig:firsteigenstateLD}
			Squared components of the first principal component ($n=1$)  of $\tilde{\bm{H}}$ for 
			(a) LDM(L), (b) LDM(H), and (c) LDM(A). }
	\end{figure}

Figure \ref{fig:LDM_svd} shows the eigenvalues $\tilde{h}_n$ for the LDM variants considered. Interestingly, 	after conditioning the LDM on binding energy data, the largest eigenvalue  $\tilde{h}_1$ dominates so much that already $S_2>99\%$ (one needs two eigenvectors to get over the  99\% threshold). This means that  there is only one direction in the space of the LDM parameters that practically matters.  To show it more explicitly, in Fig.~\ref{fig:firsteigenstateLD} the 
individual components of $\bm{V}_1$ are shown for LDM(L), LDM(H), and LDM(A).
The  LDM(H) and LDM(A) variants are strikingly close to the limit \eqref{V1}, which indicates the existence of one principal direction, which corresponds to a democratic combination of all four LDM parameter directions.
%%%%%		
	
The cumulative percentages are shown in Fig.~\ref{fig:LDM_svd}(b). One can conclude that, from a statistical perspective, the principal component analysis  of the LDM shows that $99\%$ of the variations in the data can be localized in only two linear directions of the parameter space. In that sense the model can be reduced to 2 effective parameters, for all the calibration variants considered. For all variants except LDM(L), a properly composed one-dimensional parametrization could already explain about $95\%$ of the data variability. This confirms the bivariate distributions of the LDM parameters shown in Figs. \ref{fig:LDM_bayes} and \ref{fig:LDM_bayes_A_LH} where we can see very strong posterior correlations between the parameters.

\section{Bayesian Model Averaging}\label{sec:BMA}

 Bayesian model averaging (BMA) is the natural Bayesian framework in scenarios with several competing models $\mathcal{M}_1, \dots, \mathcal{M}_K$ when one is not comfortable selecting a single model at the desired level of certainty \cite{Hoeting1999,Was00,Bernardo1994}. For any quantity of interest ${\cal O}$, e.g., the value $y^*$, the BMA posterior density $p({\cal O}|y)$ corresponds to the mixture of the posterior predictive densities of the individual models:
\begin{equation} \label{eqn:posteriorBMA}
p({\cal O}|y) 
= \sum_{k=1}^K p({\cal O}|y,\mathcal{M}_k) p(\mathcal{M}_k|y),
\end{equation}
where $y = (y_1, \dots, y_N)$ are given datapoints (here: experimental binding energies). Sampling from the BMA posterior density is trivial once one obtains posterior samples from each model. The posterior model weights $p(\mathcal{M}_k|y)$ are the posterior probabilities that a given model is the hypothetical \emph{true} model; it is  given by a simple application of Bayes' theorem:
\begin{equation} \label{eqn:posteriorsMmodel}
p(\mathcal{M}_k|y)
= \frac{p(y|\mathcal{M}_k)\pi(\mathcal{M}_k)}{\sum_{\ell=1}^K p(y|\mathcal{M}_\ell) \pi(\mathcal{M}_\ell)},
\end{equation}
where	$\pi(\mathcal{M}_k)$ are  the prior model probabilities which we choose as uniform. The so called evidence (integrals)  $p(y|\mathcal{M}_k)$ are 
obtained by integrating the data likelihood against 
the prior density of the model parameters, namely
\begin{equation} \label{eqn:evidencel}
p(y|\mathcal{M}_k) =\int p(y|\theta_k,\sigma_k,\mathcal{M}_k)\pi(\theta_k, \sigma_k|\mathcal{M}_k)d\theta_k d\sigma_k.
\end{equation}

In our study, we wish to select a model's weight according to its true predictive ability and also to avoid overfitting,
in the same spirit as the approach implemented in \cite{Neufcourt2019, Neufcourt2020,Neufcourt2020a}.
To this end, we evaluate the evidence integrals over a set of binding energies $y^*$ from the intermediate domain of Fig.~\ref{LDM_variants}, which corresponds to integrating 
the posterior distribution of new predictions 
against the posterior distribution of the model parameters
\begin{equation} \label{eqn:evidence_calc}
p(y^*|y,\mathcal{M}_k) = \int p(y^*|y,\theta_k,\sigma_k,\mathcal{M}_k)
p(\theta_k, \sigma_k|y,\mathcal{M}_k)d\theta_k d\sigma_k.
\end{equation}
Given that posterior distribution of the parameters reflects
the true distribution of the parameter more accurately than 
the prior, Eq.~\eqref{eqn:evidence_calc} more accurately represents 
the probability that $\mathcal{M}_k$ can explain data $y$.

The integral  \eqref{eqn:evidence_calc} can be transparently estimated from a convergent Markov chain as
\begin{equation}\label{eqn:evidence_mcmc}
\widehat{p(y^*|y,\mathcal{M}_k)} = \frac{1}{n_{MC}} \sum_{i = 1}^{n_{MC}} p(y^*|y,\theta^{(i)}_k, \sigma^{(i)}_k,\mathcal{M}_k),
\end{equation}
where $(\theta^{(i)}_k, \sigma^{(i)}_k)$ are samples 
from the posterior distributions $p(\theta_k, \sigma_k|y,\mathcal{M}_k)$,
which can be conveniently recycled 
from the Bayesian calibration stage (Figs. \ref{fig:LDM_bayes} and \ref{fig:LDM_bayes_A_LH}).

Evidence integrals \eqref{eqn:evidence_calc} and their estimates \eqref{eqn:evidence_mcmc} are very sensitive quantities. In general, evidences \eqref{eqn:evidence_calc} shall decrease exponentially with an increasing RMSD or a number of independent points used to compute the likelihood 
(i.e., number of evidence datapoints).  
The evidences peak at the maximum likelihood estimate of $\sigma$ but eventually fall down to zero with increasing $\sigma$. Consequently, BMA easily ends up performing a model selection instead of averaging; in practice obtaining reasonable weights requires a careful tuning of both the size of the domain on which evidence integrals are computed and the value of $\sigma$ in \eqref{eqn:stat_model}.

To assess the impact of the number of evidence datapoints, 
we evaluate evidence integrals both on the full intermediate domain $\mathcal{D_I}$ and a smaller central domain $\mathcal{D_C}$. To investigate the impact of $\sigma$, we compare the posterior weights obtained in a ``free $\sigma$'' setup described in Sec.~\ref{BAnal} where $\sigma$ is determined by its posterior distribution guided by the data, with these obtained taking $\sigma$ fixed to an \emph{a priori} value same for all LDM variants (L, H, H+L, A). 
When $\sigma$ is fixed, 
$\sigma_k$ and $\sigma^{(i)}_k$
can simply be ignored in 
Eqs.~\eqref{eqn:evidencel}-\eqref{eqn:evidence_mcmc},
and set to the fixed value $\sigma^{(i)}_k := \sigma$ 
in Eq. \eqref{eqn:evidence_laplace}.
While the free-$\sigma$ variant is more natural and ``honest'', it lets $\sigma$ drift towards the points associated with larger residuals, which reduces the difference between model evidences. The fixed-$\sigma$ variant allows to control for the impact of $\sigma$ on the weights of the models constrained on different domains.

The extreme numerics of likelihoods can sometimes take us close to machine limits since a part of the Monte Carlo likelihood samples in Eq.~\eqref{eqn:evidence_mcmc} can fall below the double precision floating point. One can mitigate this problem by discarding all the likelihood samples for which $p(y^*|y,\theta^{(i)}_k,\sigma^{(i)}_k, \mathcal{M}_k)$ is evaluated to be a numerical zero or by rescaling the evidences by an arbitrary common factor.
Since the evidence estimator is a simple average, it is also extremely sensitive to outliers and one large value of $p(y^*|y,\theta^{(i)}_k,\sigma^{(i)}_k,\mathcal{M}_k)$ can outweigh all the remaining samples; 
we consider as outliers these likelihood samples 
falling behind 3-sigma intervals and discard those. 
In view of these instabilities, for comparison
we  also compute
the pseudo-evidence
\begin{equation} \label{eqn:evidence_laplace} 
p(y^*|y,\widehat{\theta}_k, \widehat{\sigma}_k,\mathcal{M}_k)
:=  \frac{1}{(\sqrt{2\pi}\widehat{\sigma}_k)^{n^*}} 
\exp{\left(-\frac{1}{2\widehat{\sigma}_k^2}\sum_j(y^*_j-f(x^*_i,\hat{\theta}))^2\right)}, 
\end{equation} 
where $x^*_j$ are the locations of $y_j^*$, $j=1, ..., n^*$, and  $(\widehat{\theta}_k, \widehat{\sigma}_k)$ are the posterior means of the parameters ($\widehat{\sigma}_k = \sigma$ in fixed-$\sigma$ case); this corresponds to replacing
the posterior distribution
$p(\theta_k, \sigma_k|y,\mathcal{M}_k)$ in Eq.~\eqref{eqn:evidence_calc}
by a Dirac delta function at its (posterior) mean. 
The resulting quantity can be thought of as the counterpart 
for the evidence \eqref{eqn:evidence_calc}
as is the Laplace approximation for classical BMA factors based on the training data;
for the evidence integral \eqref{eqn:evidencel},
the Laplace approximation consists in
a second order Taylor approximation of the logarithm of the likelihood in \eqref{eqn:evidencel} around the maximum of the posterior distribution $p(\theta_k,\sigma_k|y,\mathcal{M}_k)$. In this way, the log-likelihood becomes Gaussian and Eq.~\eqref{eqn:evidencel} has a closed-form expression \cite{KassRaftery1995}. The Laplace method works well for very peaked likelihoods. We can illustrate the underlining idea behind Eq.~\eqref{eqn:evidence_laplace} 
by considering two simple limiting cases:
	\begin{description}
		\item[i]
		All models are similar in the sense that
		the posterior predictions have similar average deviations: the posterior weights 
		in both \eqref{eqn:evidence_calc} and \eqref{eqn:evidence_laplace} are the same;
		\item[ii]
		One model is much better than the others, in the sense that its posterior predictions have a higher likelihood: this model
		attains a weight close to 1 in both \eqref{eqn:evidence_calc} and \eqref{eqn:evidence_laplace}.
	\end{description}

By using BMA to combine  models, we are accounting for an additional source of uncertainty that is not considered by individual models. In fact, \cite{Kej19} showed that the mean of the BMA posterior density \eqref{eqn:posteriorBMA} leads to more accurate predictions and can improve the fidelity of the posterior credible intervals from individual models. 
However, we wish to emphasize that the definition of BMA relies 
on the assumption that the data distribution actually follows 
one of the models and the model set is complete. This is not always the case, especially in the context of nuclear modeling, and one may need to relax this assumption in practice (as we did in the LDM study here). This is a clear limitation to the suitability of BMA to combine several imperfect models and consequently make the parameter $\sigma$ a key player in the calculation of the evidence and the ranking of models: a model with a larger $\sigma$ is weaker in the sense that it contains less information and less commitment -- it tends to yield larger evidence and larger model weight; on the contrary a model with a small $\sigma$ is more likely to be proved wrong by the data and to be attributed a lower weight. On a similar note, a lower $\sigma$ implies a lower tolerance to discrepancies, while a $\sigma$ large enough can tolerate discrepancies as large as desired.

%%%%%%%%%%%%%%%%%%%%%%%%%%%%%%%%%%%%%%%%%%%%%%%%%%%%%%
\subsection{Results}
%%%%%%%%%%%%%%%%%%%%%%%%%%%%%%%%%%%%%%%%%%%%%%%%%%%%%%

\begin{figure}[htb!]
	\center
	\includegraphics[width=0.8\linewidth]{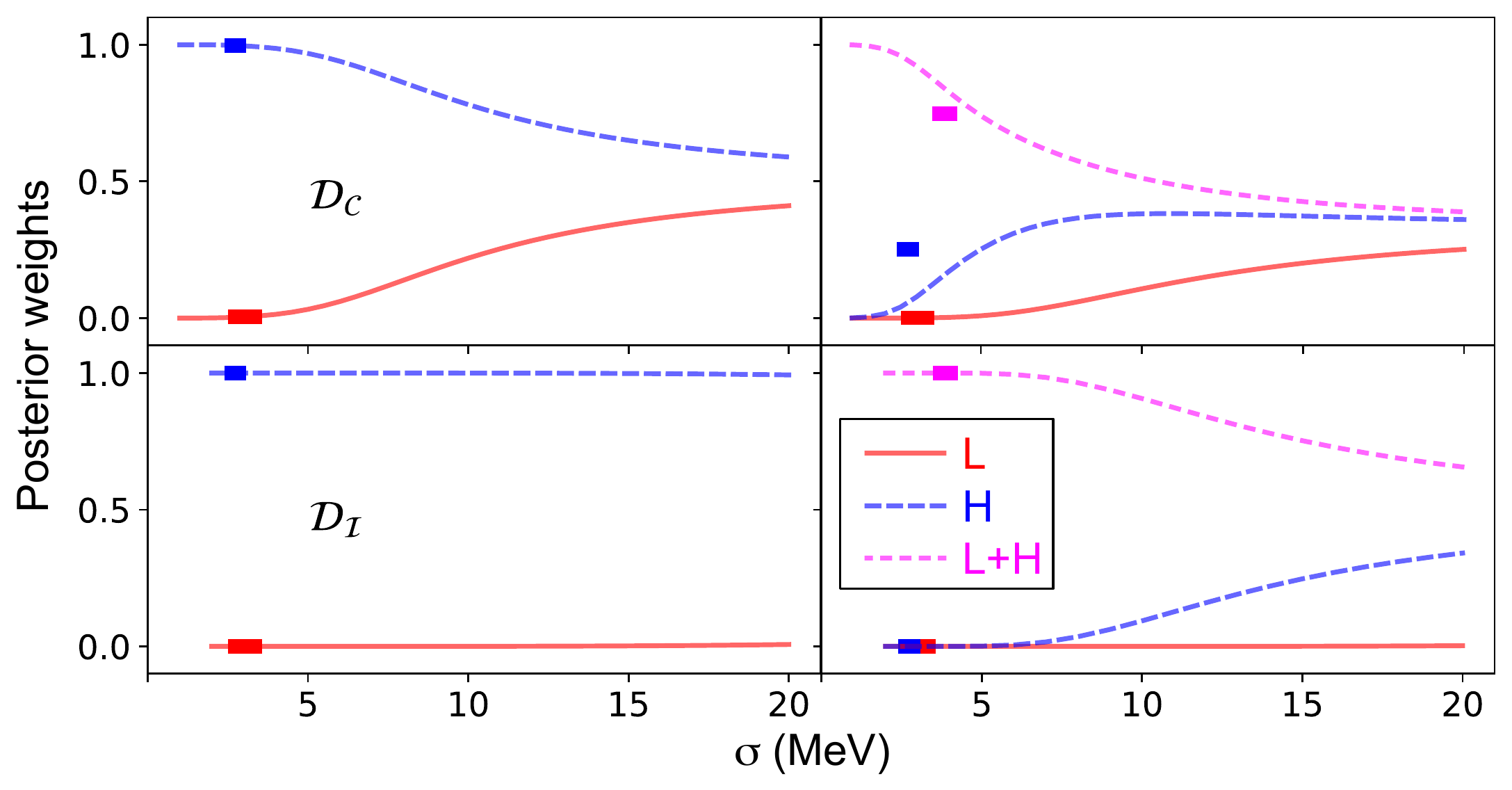}
	\caption{\label{fig:BMA_weights}
Posterior model weights under the averaging scenarios with two (L and H; left) and three  (L, H, and L+H; right) models. The model weights in the fixed-$\sigma$ setup are shown by lines. The boxes mark the model weights (ordinate) and 
$99.7\%$ HPD credible intervals (abscissa) 
-- the Bayesian counterpart to frequentist 3-sigma confidence intervals -- 
for $\sigma$ in the free-$\sigma$ setup. 
Evidences are evaluated on subset $\mathcal{D}_\mathcal{C}$ of 8 nuclei in the intermediate domain (top panels) and on the full intermediate domain $\mathcal{D}_\mathcal{I}$ of 155 nuclei (bottom panels).
}
\end{figure}

Figure~\ref{fig:BMA_weights} shows the posterior weights obtained in the two- (left) and three- (right) model variants.
We compare several setups with the
evidence integrals  computed both 
on the full intermediate domain $\mathcal{D}_\mathcal{I}$ 
and on a smaller subset of nuclei $\mathcal{D}_\mathcal{C}$.
Scaling $\sigma$ is taken either as a fixed
or free parameter. The corresponding RMSD values are listed in Table~\ref{tab:rms} (denoted BMA).
%%%

As expected, model (H) is selected in the two model variant, 
and the (L+H) variant dominates when it is included -- 
this is true for both the free-$\sigma$ variant and the fixed-$\sigma$ variant, and for both sets of evidence datasets $\mathcal{D_C}$ and $\mathcal{D_I}$. 
This is consistent with the RMSD of these models. 
It shall be emphasized that BMA performs a model selection 
in the two-model variant, where the RMSDs of 
the competing models are very different,
and proper model averaging in the three-model variant,
where the RMSD of (H) and (L+H) are close enough.
Table \ref{tab:rms} also shows how the RMSD 
of the BMA predictions compare with that of individual models. In the two-model setup, BMA is very much like (H) and it has a similar RMSD. In the three-model setup, BMA performs much better than the worst model and very close to the best of the averaged models. 
When computed on the full test domain $\mathcal{D_I}$, RMSDs are systematically smaller for the BMA than for all the individual models involved in the averaging (not considering LDM(A)). One may notice that the RMSD of BMA(L, H, L+H) is, perhaps unexpectedly, slightly worse than that of LDM(L+H) on the small domain $\mathcal{D_C}$. However, these values are based merely on 8 data points and should be viewed as  a crude estimate of true predictive performance.

\begin{figure}[htb!]
	\center
	\includegraphics[width=0.8\linewidth]{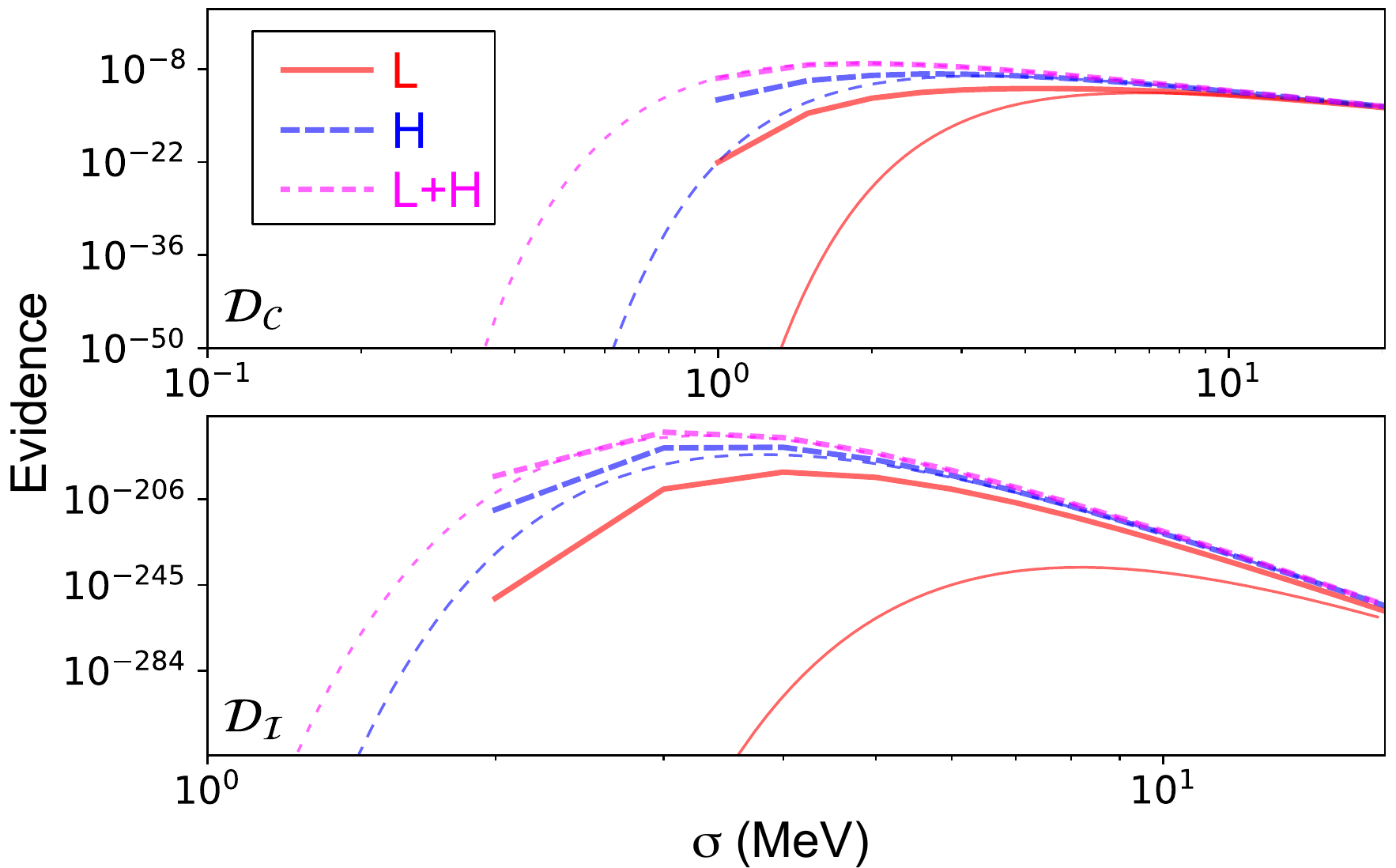}
	\caption{\label{fig:LDM_evidence_log}
		The evidences calculated from MCMC samples (thick lines)
		and the Laplace approximations at the posterior mean $\hat\theta$ (thin lines) in the scenario with fixed $\sigma$ for   $\mathcal{D}_\mathcal{C}$ (top) and $\mathcal{D}_\mathcal{I}$ (bottom).
	}
\end{figure}

We investigate further the posterior weights in Fig.~\ref{fig:LDM_evidence_log} by comparing directly the evidences obtained for the three variants (L), (H), (L+H) for fixed $\sigma$. We also show the approximations \eqref{eqn:evidence_laplace} of the evidences at the posterior mean value of the LDM parameters, again for fixed $\sigma$. We see that evidences are very small and  quickly approach zero at low and large  values of $\sigma$; the right tail is linear in the log space, with the slope approximately given by the number of datapoints. 
%Note that the leftmost points in Fig.~\ref{fig:LDM_evidence_log} are below the double floating-point precision; hence, they are not shown.

As  discussed above, we investigate the impact of $\sigma$ values on the evidence integrals by comparing the posterior weights when $\sigma$ is fixed and when it is considered a free parameter. In the fixed-$\sigma$ variant, we expect that the posterior weights converge to the prior weights when $\sigma\to\infty$: in this limiting case, all RMSDs are relatively small and corresponding evidences  go to zero. On the contrary, at the 
small-$\sigma$ limit, the model with the lowest RMSD receives a weight of 1. This is clearly seen in Fig.~\ref{fig:BMA_weights}: the best model is selected with a weight of 1 at a low $\sigma$, and the weights progressively converge to the uniform priors. The convergence speed towards uniform weights increases with the closeness of the RMSDs between the models 
and the number of data used in the evidence evaluation 
(number of evidence datapoints).

When it comes to fixing $\sigma$ to an arbitrary value, one needs to be particularly cautious due to the numerical difficulties related to a likelihood computation. Consequently, the scaling parameter $\sigma$ must be carefully chosen to be in the domain where the numerical values produced are meaningful. If it is not clear \emph{a priori} what the value needs to be taken for $\sigma$, we see two reasonable approaches. The first is to select the value at which the evidence (or its approximation around RMSD) is maximized. This should be close to its maximum likelihood estimate. A preferable option is to take $\sigma$ as determined by the data, i.e., taken under its posterior distribution.

In Fig.~\ref{fig:LDM_evidence_log} we compare
the evidence calculated from MCMC samples and the Laplace approximations at the posterior mean of $\theta$.
Recall that we are computing the evidence integrals as \eqref{eqn:evidence_calc}, 
thus used the samples directly from the posteriors in \eqref{eqn:evidence_mcmc}. 
These samples are reasonably centered around the posterior means,
so the integrals should be reasonably close to their 
Laplace approximates at the posterior parameter means,
namely \eqref{eqn:evidence_laplace}.
Therefore the impact of the non-zero width of the 
posterior distribution of $\theta$ shall be limited.
While the agreement is very good for the (H) and (L+H) models,
 we observe an important difference for model (L).
In light of the large RMSD of (L) 
and its relatively low $\sigma$ values,
this could be explained by (L) having
posterior parameter distributions localized 
too far from the maximum
of the likelihood.
In general, these discrepancies between the MCMC estimates of the evidences
and Laplace approximations are not unexpected 
(see \cite{KassRaftery1995,Hoeting1999})
and can be attributed to the combination of approximations inherent in MCMC methods and the nature of Laplace approximation.

When comparing the results obtained on the two integration domains, we also see that the length of the $\sigma$ domain on which the weights transition is sharper when the domain is smaller. Both Figs. \ref{fig:BMA_weights} and \ref{fig:LDM_evidence_log} clearly illustrate that it is easier to compute evidences on a smaller domain and impractical to use a large domain to obtain meaningful averaging.
Nevertheless Laplace approximation continues to produce 
sensible estimates for the evidences on a large number of points,
which are calculable at the logarithmic scale
and thus more robust to numerical issues.

\begin{figure}[htb!]
	\center
	\includegraphics[width=0.8\linewidth]{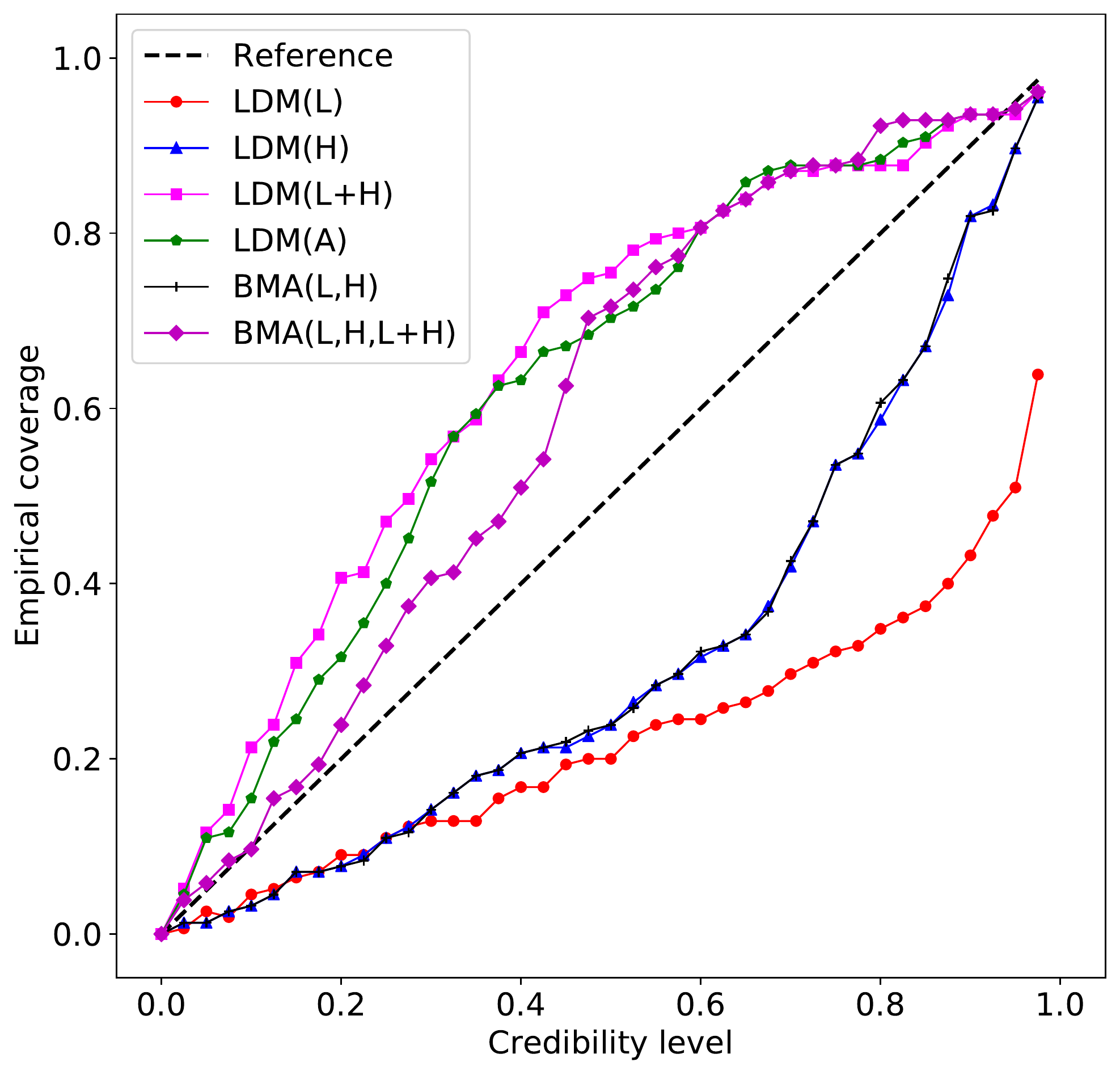}
	\caption{\label{fig:LDM_ECP}
		Empirical coverage probability for the four LDM variants used in our study and the averaging scenarios with two (L and H) and three models (L, H, and L+H). The empirical coverage was calculated based on HPD credibility intervals. 
	}
\end{figure}

\subsection{Empirical coverage probability}

In addition to evaluating the BMA from the prediction accuracy point of view, we present in Fig. \ref{fig:LDM_ECP} what is know as the empirical coverage probability (ECP)  \cite{Gneiting2007a,Gneiting2007}. 
The ECP is an intuitive approach to measuring the quality of a statistical model's UQ. Formally, it can be written as
\begin{equation}
\eta(t):=\frac{1}{n}\sum_{i=1}^n 1_{y_i^* \in I_t(x_i^*)},
\end{equation}
where $1$ is the indicator function (which is 1 when the 
argument is true and 0 otherwise), 
$I_t(x_i^*)$ is the $t-$credibility interval produced by the calibrated model at a new input $x_i^*$,  and $y_i^*$ are the (new) testing data.

Each line in Fig. \ref{fig:LDM_ECP} represents the proportion of a model's prediction of independent testing points falling into the respective credibility intervals (highest posterior density intervals). These lines should theoretically follow the diagonal so that the actual fidelity of the interval corresponds to the nominal value. If the respective ECP line falls above the reference, credible intervals produced by a given model are too wide (UQ is conservative). Naturally, a model with an ECP line below the reference underestimates the uncertainty of predictions (UQ is liberal). While the values of empirical proportions close to the reference curve are desirable, it is usually preferable to be conservative rather than liberal. Overly narrow credible intervals declare a level of assurance higher than it should be. 

Figure~\ref{fig:LDM_ECP} shows that the LDM variants fitted to the smaller domains (L or H) tend to underestimate the uncertainty of the predicted binding energies compared to the rather conservative UQ of the (L+H) variant and the LDM fitted to the entire AME2003 dataset. 
There is an interesting comparison to be made between the ECP curves and the posterior distributions of $\sigma$ in Fig.~\ref{fig:LDM_bayes_sigma}. The posterior means of $\sigma$ in the  LDM(L)  and LDM(H)  variants are significantly smaller than those of LDM(L+H) and LDM(A), which consequently makes these models too liberal in their UQ. Note that it is not surprising that the ECP for BMA of LDM(L+H) coincides with the ECP of LDM(H)  since the model weight is 1 for all practical purposes. On the other hand,  BMA(L,H,L+H) yields an ECP superior to all the LDM variants, including LDM(A), which aligns with our hypothesis that meaningful averaging can lead to an improved UQ.

	\section{Realistic DFT calculations}
	
In this section,  we investigate the structure of the realistic Skyrme energy density functional used in self-consistent DFT calculations of nuclear masses. 
We first apply the chi-square correlation technique to  study different Skyrme models obtained by model calibration using homogeneous and  heterogeneous datasets. We then carry out the principal-component  analysis  to learn about the number of effective degrees of freedom of the Skyrme functional.

	\subsection{The Skyrme functional}\label{Skyrmef}

	As a microscopic alternative to the LDM, we investigate the
	Skyrme-Hartree-Fock (SHF) model, which is a widely used representative
	of nuclear density-functional theory
	\cite{Ben03,Erl11aR}. The SHF model aims at a self-consistent description of
	nuclei, including their bulk properties  and  shell
	structure. We summarize here briefly the Skyrme energy functional
	which is used for computing time-even ground states. It is formulated
	in terms of local nucleonic  densities: particle density $\rho$, kinetic density
	$\tau$, and spin-orbit density $\vec{J}$.  The Skyrme energy density can be written as
	\begin{eqnarray}
	\hspace*{-2em}
	{\cal E}_{\rm Sk}
	&=&
	\underbrace{
		\frac{1}{2}b_0\rho^2+\frac{1}{2}b'_0\sum_t\rho_t^2
		+
		\frac{1}{6}\rho^\alpha\left(b_3\rho^2+\frac{1}{6}b'_3\sum_t\rho_t^2\right)
		+
		b_1\rho\tau+b'_1\sum_t\rho_t\tau_t
	}_{\mathrm{bulk}}
	\nonumber\\
	&&+
	\underbrace{
		\frac{1}{2}b_2\rho\Delta\rho+\frac{1}{2}b'_2\sum_t\rho_t\Delta\rho_t
	}_{\mathrm{surface}}
	+
	\underbrace{
		b_4\rho\nabla\!\cdot\!\vec{J}+b'_4\sum_t\rho_t\nabla\!\cdot\!\vec{J}_t
	}_{\mathrm{spin-orbit}}
	\;,
	\label{eq:basfunct}
	\end{eqnarray}
	where $t\in\{\mbox{proton,neutron}\}$ and the total density is the sum of proton and neutron contributions.
	The energy density (\ref{eq:basfunct}) has 11 free parameters, the 10 $b$ parameters and the exponent $\alpha$.  A density-dependent pairing functional
	is added, which is characterized by three parameters: the pairing strengths
	$V_{\mathrm{pair},t}$ and
	reference density $\rho_{0,\mathrm{pair}}$. These 14 parameters are adjusted by least-squares fits \cite{Kluepfel2009} to deliver a global description
	of all nuclei, except for the very light ones. The model parameters  can be  sorted into three groups: bulk, surface, and spin-orbit. Bulk
	properties can be equivalently expressed by the symmetric nuclear matter
	parameters (NMP) at equilibrium: binding energy per nucleon  $E/A$, saturation density
	$\rho_0$, compressibility $K$, symmetry energy $J$,
	symmetry energy slope $L$, isoscalar effective mass $m^*/m$, and isovector
	effective mass expressed in terms of the  sum rule enhancement
	$\kappa$. Bulk surface properties can also be expressed in terms of surface
	energy $a_\mathrm{s}$ and surface-symmetry energy $a_\mathrm{ssym}$.	
	 Most of these parameters
	can be related to those of the  LDM \cite{Reinhard2006}. It is only effective mass and spin-orbit parameters that
	are specific to shell structure and go beyond the LDM. Experience
	shows that a definition of the Skyrme functional through NMP is better
	behaved in least-squares optimization which indicates that a physical
	definition is superior over a technical definition \cite{UNEDF0}. We shall return to this point later.
	
	The original formulation of the SHF method was
	based on the concept of an effective density-dependent interaction, coined the Skyrme
	force \cite{Sky59a}, which was used to derive the density functional
	as expectation value over a product state  $ |\Phi\rangle$:
	\begin{eqnarray}
	\hspace*{-2em}
	{\cal E}_{\rm Sk}^{\mathrm{int}}
	&=&
	\langle\Phi|
	t_0(1\!+\!x_0 \hat{ P}_\sigma)\delta({\boldsymbol r}_{12})
	+ \frac{t_3}{6}(1\!+\!x_3\hat P_\sigma)\rho^\alpha\left({\boldsymbol r}_1\right)
	\delta({\boldsymbol r}_{12})
	\nonumber\\
	&&
	+\frac{t_1}{2}(1\!+\!x_1\hat{P}_\sigma)
	\left(
	\delta({\boldsymbol r}_{12})\hat{\boldsymbol k}^2
	+
	{\hat{\boldsymbol{k}}}^{2}\delta({\boldsymbol r}_{12})
	\right)
	+ t_2(1\!+\!x_2\hat P_\sigma)\hat{\boldsymbol k}
	\delta({\boldsymbol r}_{12})\hat{\boldsymbol k}
	|\Phi\rangle,
	\label{eq:skenfun}
	\end{eqnarray}
	where ${\boldsymbol r}_{12}={\boldsymbol r}_1-{\boldsymbol r}_2$, $\hat{P}_\sigma=
	\frac{1}{2}(1+\hat{\boldsymbol \sigma}_1\hat{\boldsymbol{\sigma}}_2)$
	is the spin-exchange operator, and $\hat{\boldsymbol k}$ is the
	momentum operator.  The model parameters 
	of (\ref{eq:skenfun}) are ($t_i$, $x_i$, $\alpha$). These 11 parameters are
	fully equivalent to the above 11 SHF parameters (7  NMP plus 2 surface and 2
	spin-orbit parameters). But the degree of correlation among parameters can
	be very different as we shall see below. 
	
	In this study, we shall primarily  use  two Skyrme functionals: 
	SV-min \cite{Kluepfel2009} and SV-E (a simplified version of functional E-only of Ref.~\cite{Erler2015}). These two functionals differ in their datasets  of fit-observables. The  basic dataset of SV-min  \cite{Kluepfel2009} 
	contains selected experimental data on binding energies, charge radii,
	diffraction radii, surface thickness, pairing gaps deduced from odd-even binding energy staggering, and spin-orbit splitting. 
	The model SV-E is introduced  to check the impact of the fit-data; it has been solely informed by the binding-energy subset of the SV-min dataset. Recall that
	the LDM is also  fitted exclusively to
	binding energies. 
	
	The remaining functionals used in this work  are SV-min(t,x)  and SV-bas. SV-min(t,x)  is the same model as SV-min but expressed in
	terms of the original Skyrme parameters $(t_i,x_i,\alpha$) rather than NMP. 
	In order to clearly distinguish between  these two parametrizations, we shall 
	use the alternative name  SV-min(NMP) for SV-min.
	The functional SV-bas has been optimized to the dataset of SV-min
	augmented by the data  from four giant resonances \cite{Kluepfel2009}.

	\subsection{Correlation analysis}

	The further processing of the Skyrme model is the
	same as from the LDM above, 
	starting with parameter
	optimization by minimizing the penalty function, probabilistic
	interpretation, and subsequent principal component analysis of the emerging
	Hessian matrix. There is only one important difference in
	the design of the penalty function. The form in \eqref{eqn:chisq} requires that all observables $y_i$ be of the same nature and have the same dimensions. This is also why the penalty function \eqref{eqn:chisq} does not depend on the scale $\sigma$. The fit to the dataset
	of \cite{Kluepfel2009}, however, includes different kinds of observables
	(energies, radii, ...) and associates to them different
	weights in the composition of the penalty function, now
	reading
	\begin{equation}
	\chi^2(\theta)
	=
	\sum_{i=1}^N\frac{\left(y_i - f(x_i, \theta) \right)^2}{\sigma_i^2}.
	\label{eq:chi1}
	\end{equation}
	Everything else, the Hessian and its handling, remains as outlined above.

	In the language of CoDs, the presence of parameter correlations means that the conditioned covariance matrix
	$\tilde{\bm{C}}$ has a considerable amount of non-diagonal entries,
	and the same holds for the conditioned Hessian 
	$\tilde{\bm{H}}$. Both matrices  have
	diagonal elements one throughout and $\det(\tilde{\bm{H}})\le 1$. In fact, this determinant
	can become very small in large parameter spaces often driving the
	linear algebra toward the precision limit. In the worst case, the
	determinant of the Hessian becomes zero; hence, its covariance matrix
	is singular. Such a situation can be handled with the help of a
	SVD technique, see Sec.~\ref{SVD-Sk}. 
	
		\begin{figure}[htb!]
		\center
		\includegraphics[width=0.7\linewidth]{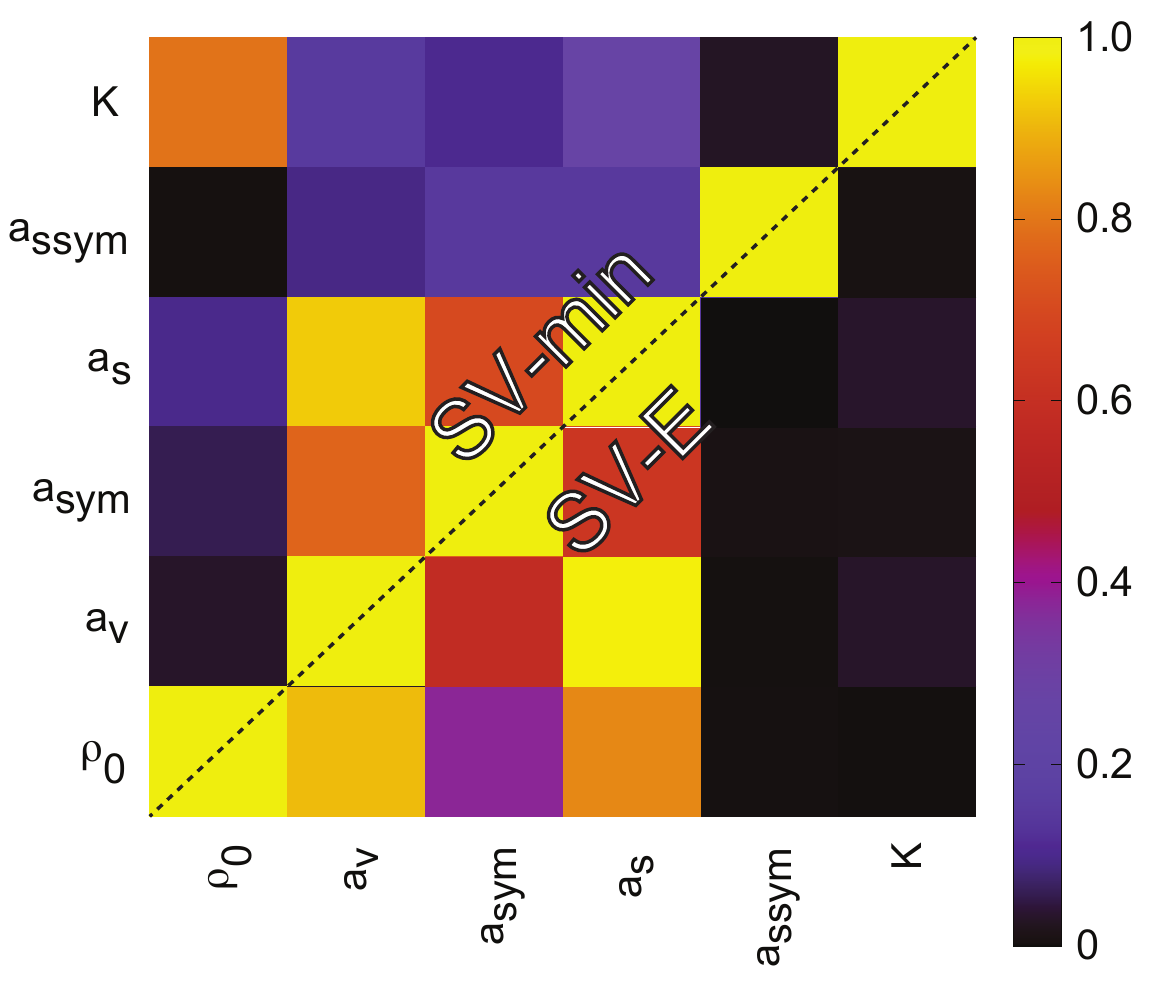}
		\caption{\label{SVcorr}
			Matrix of CoD between the subset of Skyrme model parameters 
			related to LDM parameters: equilibrium density $\rho_0$, volume energy
			$a_{\rm v}\equiv E/A$, (volume) symmetry energy $a_\mathrm{sym}\equiv J$, surface
			energy $a_\mathrm{s}$, surface symmetry energy $a_\mathrm{ssym}$, and
			compressibility $K$. Two matrices are shown, the upper triangle
			for SV-min \cite{Kluepfel2009} and the lower triangle for SV-E which
			is fitted to the subset of the fit-data  from \cite{Kluepfel2009}
			involving only binding energies.
		}
	\end{figure}
	%%%%%%%%%%%%%%%%%%%%%%%%%%
	Figure~\ref{SVcorr} shows
	the matrix of CoD between the LDM subset of Skyrme model parameters
	for SV-min  and SV-E (cf. also Fig. 8 of Ref.~\cite{Erler2015}). Although different
	in details, both parametrizations produce a considerable amount of
	correlations between $a_{\rm v}, a_{\rm sym}$, and $a_{\rm s}$. 
	The saturation density $\rho_0$ is not correlated with these LDM parameters for SV-min. Indeed, in this case $\rho_0$ is primarily constrained by the data on charge and diffraction radii \cite{Reinhard2016}. Since the radial information is missing in the dataset of SV-E, appreciable correlations between
	$\rho_0$  and ($a_{\rm v}, a_{\rm s}$) appear, as in Fig.~\ref{fig:LDMcorr} for the LDM case. This indicates
          that more data can reduce parameter correlations thus
          rendering more model parameters significant (see also
          Sec.~\ref{SVD-Sk}).
	
	Strong correlations between certain model parameters suggest that the
	actual numbers of model degrees of freedom (conditioned on a given dataset) is less than the number of Skyrme model
	parameters  suggests. This point will  be addressed  in the following section.

	\subsection{Principal component  analysis of the Skyrme functional}\label{SVD-Sk}
	
	The Skyrme functional described in Sec.~\ref{Skyrmef} has 14 parameters. However, as the correlation analysis indicates, some of the parameters are correlated. This raises the question of  the effective number of parameters characterizing the   Skyrme model, given the dataset of fit-observables. In practice, a more meaningful question is that of the minimum number of principal directions in the model's parameter space that are constrained by the dataset employed. Some investigations along those lines have already been carried out in Refs. \cite{Bertsch2005,Toivanen2008} in the context of Skyrme models and in 
Refs.~\cite{Niksic2016,Niksic2017,Taninah2020} in the framework of covariant density functional theory.
	
	\begin{figure}[b!]
		\centerline{\includegraphics[width=\linewidth]{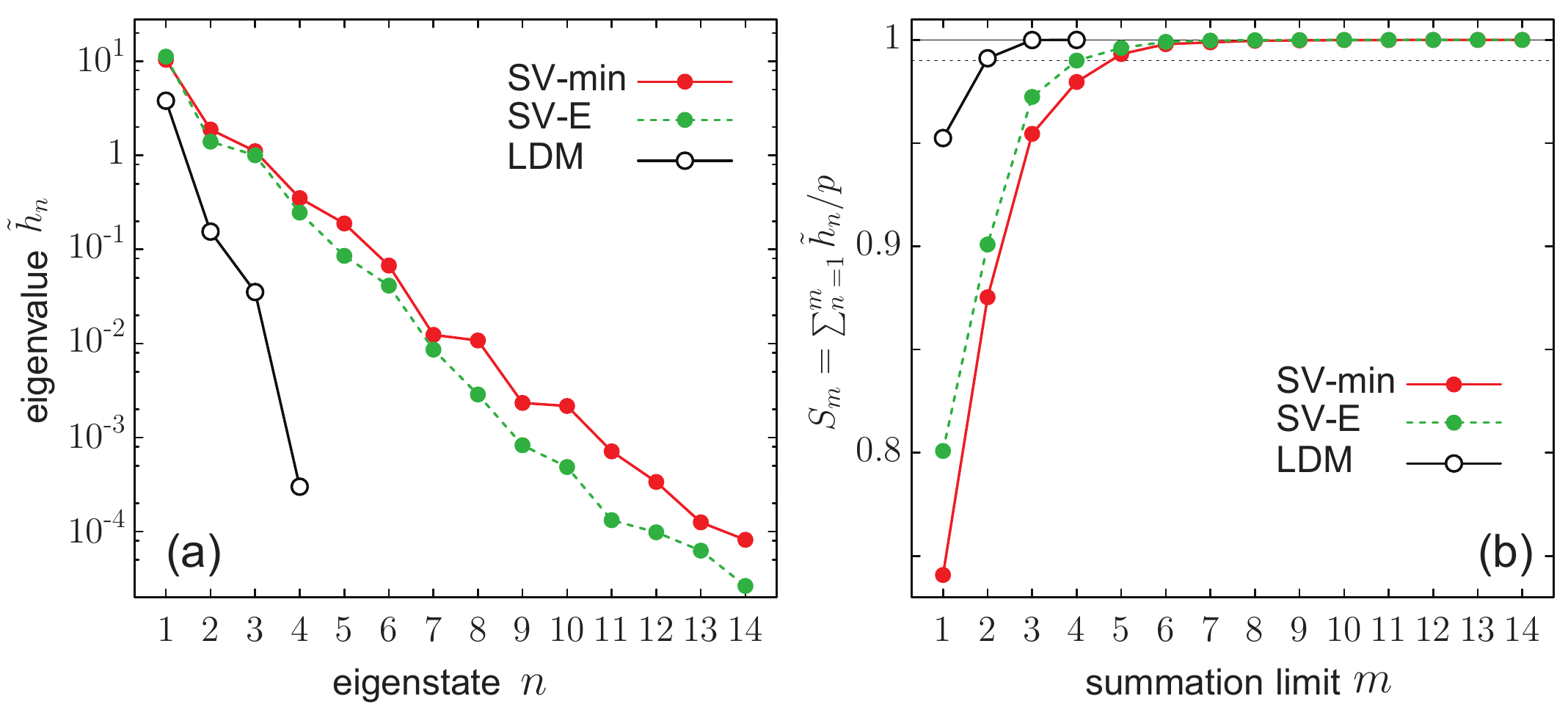}}
		\caption{\label{fig:cutoff-test}
			(a) Eigenvalues $\tilde{h}_n$ of the conditioned Hessian matrix 
			and (b) cumulative percentage (\ref{eq:exhaust})
			for the 14-parameter Skyrme functionals 
			SV-min and 
			SV-E and for the 4-parameter LDM (cf. Fig.~\ref{fig:LDM_svd} for a detailed LDM discussion). A dotted horizontal line in (b) indicates the
			99\% threshold.
		}
	\end{figure}
	%%%%
	Fig. \ref{fig:cutoff-test}(a) shows the eigenvalues $\tilde{h}_n$ 
	for SV-min, SV-E, and the LDM.
	They decrease  nearly exponentially with $n$ spanning 5-6 orders of magnitude.
	This huge range  indicates also the minimum number of digits required
	for the model parameters and the precision of observables to make a
	meaningful analysis. It is
	interesting to note the differences between the Skyrme models. SV-E, solely informed by binding energies, is less constrained by the data than SV-min.
For the  4-parameter LDM, the eigenvalues decrease very fast. This indicates a lot of redundancy in this sparse model.
	The percentage of
	the partial summation accounted for by the lowest principal components is displayed in Fig.~\ref{fig:cutoff-test}(b).  
	The highest eigenvalue $\tilde{h}_1$ exhausts from 74\% (SV-min) to 95\% (LDM) of the sum rule (\ref{eq:exhaust}) indicating a very high level of parameter correlation. 
	
Taking the reference threshold as $S_\mathrm{limit}=0.99$ reduces the number of significant parameter directions
	dramatically. For the standard Skyrme model the parameter space is reduced from
	15 to 4-5 effective parameters and for the LDM from 4 to 1. This finding is consistent with the discussion in Refs.~\cite{Bertsch2005,Toivanen2008}.
	With that result at hand, we can define an equivalent cutoff in the
	space of eigenvalues which would then come around $\tilde{h}_n=0.2$ to yield the same number of  effective parameters.

	\begin{figure}[htb!]
		\centerline{\includegraphics[width=0.8\linewidth]{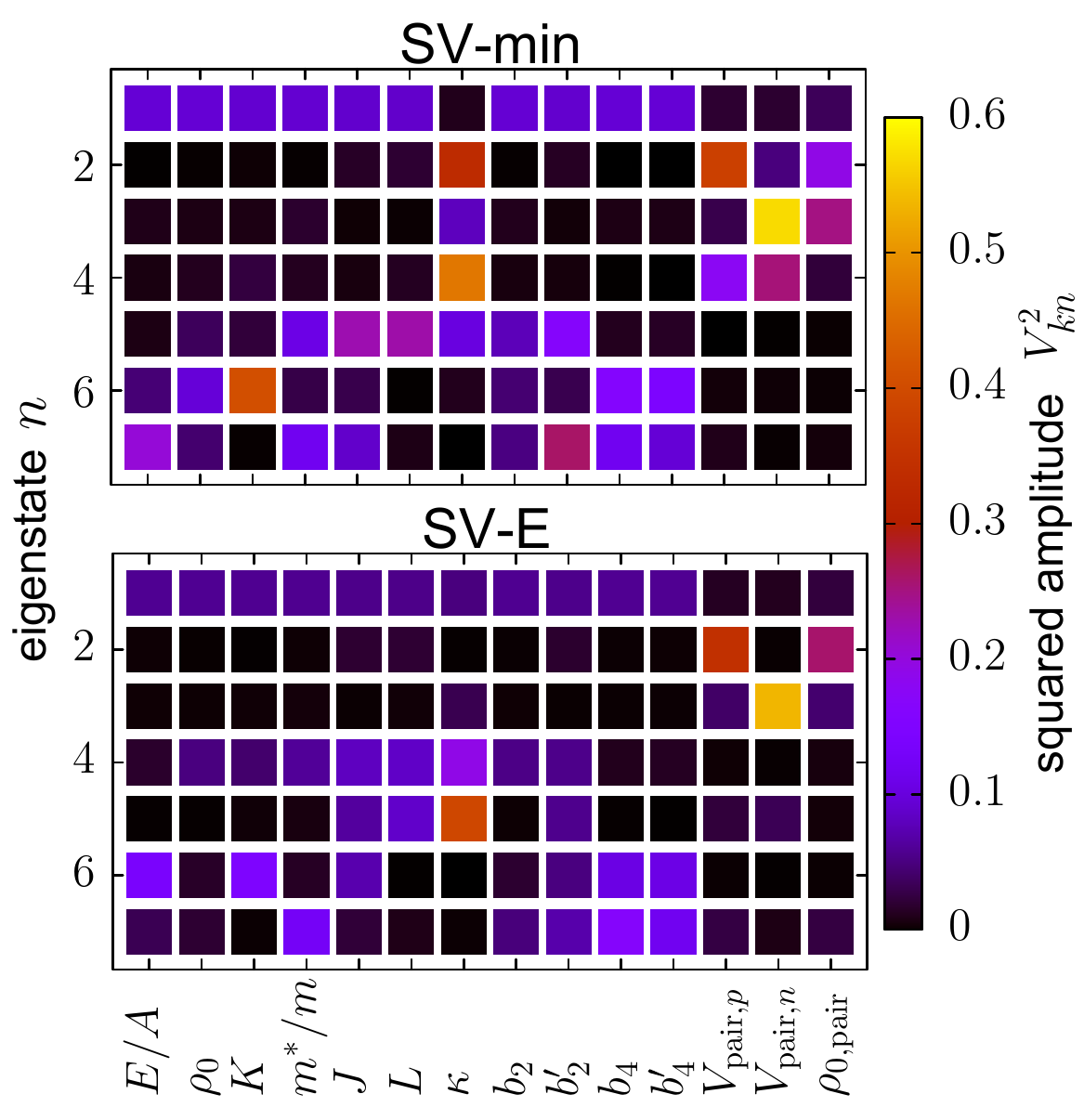}}
		\caption{\label{fig:eigenstates-SV}
			Principal components (\ref{princ}) of SV-min (top) and
			SV-E (bottom) corresponding to the 7 highest eigenvalues $\tilde{h}_n$ shown in Fig.~\ref{fig:cutoff-test}(a).  The squared amplitudes $V_{kn}^2$ are
			represented by color.}
	\end{figure}
	Fig. \ref{fig:eigenstates-SV} illustrates the composition of the
	principal components of the Hessian matrix for SV-E and SV-min.   
	For SV-min, the first four principal components primarily reside in three subspaces:  10-parameter space
	$\vartheta_1:=\{E/A, \rho_0, K, m^*/m, J,  L, b_2, b'_2, b_4, b'_4\}$;  1-parameter space $\vartheta_2:=\{ \kappa\}$; and  3-parameter space
	$\vartheta_3:= \{V_{\mathrm{pair},p}, V_{\mathrm{pair},p}, \rho_{0,\mathrm{pair}}\}$. The subspace $\vartheta_1$ is
          represented by the first principle component $n=1$; it
          consists of 10 out of the 11 parameters of the Skyrme
          functional, except $\kappa$. The third group $\vartheta_3$ (spanned by the eigenvectors $n=2-4$) consists of pairing parameters.  Surprisingly, the directions of $\vartheta_2$ and $\vartheta_3$ are slightly coupled.
	
	For SV-E, the subspaces $\vartheta_1$ and $\vartheta_3$ are very well separated, and the coupling between $ \kappa$ and the pairing subspace $\vartheta_3$ becomes vanishingly small. For both models, the isovector effective mass (quantified by $\kappa$) is very poorly constrained by the data. Indeed, the results of chi-square optimization for $\kappa$ are: $-0.18(27)$ (SV-min) and $0.10(33)$ (SV-E).
	
	\begin{figure}[htb!]
		\centerline{\includegraphics[width=0.8\linewidth]{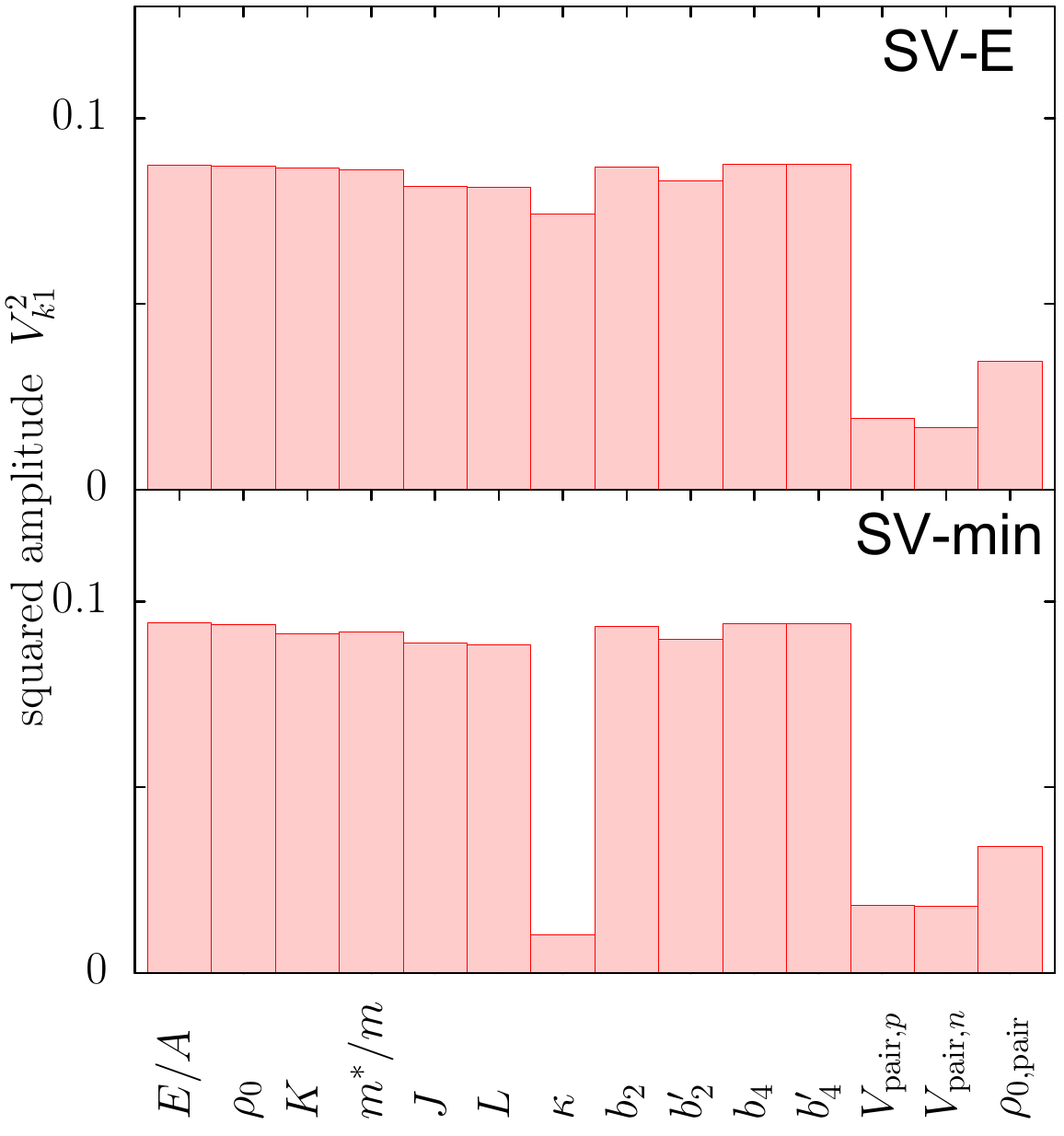}}
		\caption{\label{fig:firsteigenstate}
			Squared components of the first principal component ($n=1$)  of $\tilde{\bm{H}}$ for 
			SV-E (top) and SV-min (bottom). }
	\end{figure}
	
	The structure of the first principal component (the largest-eigenvalue eigenstate of the conditioned Hessian matrix $\tilde{\bm{H}}$) is displayed in Fig.~\ref{fig:firsteigenstate}. This plot nicely demonstrates the separation between
	the particle-hole and the pairing parameter space. What is quite remarkable is that the amplitudes of all parameters belonging to the  $\vartheta_1$ space in the case of SV-min, and $\vartheta_1 \oplus \vartheta_2$ space in the case of SV-E are virtually identical,
	$V^2_{k1}\approx 0.09$. Consequently, in these subspaces, the structure of the first principal component 
is reminiscent of
	the LDM case discussed in Fig.~\ref{fig:firsteigenstateLD} showing a very high correlation between model parameters.

	\begin{figure}
		\centerline{\includegraphics[width=\linewidth]{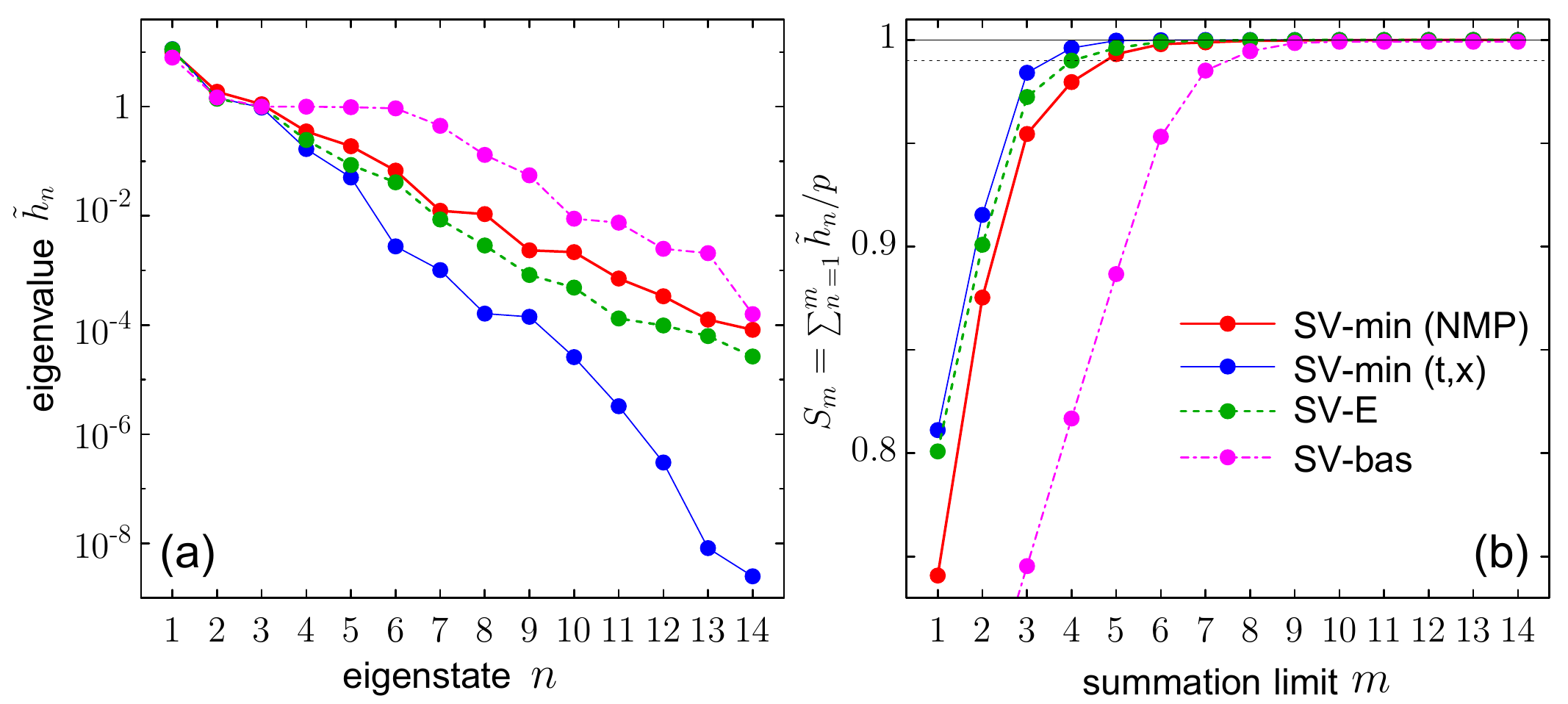}}
		\caption{\label{fig:cutoff-SVseries}
			Similar as in Fig.~\ref{fig:cutoff-test} but for the functionals SV-min(NM), SV-min(t,x), SV-E, and SV-bas.
		}
	\end{figure}

	Figure~\ref{fig:cutoff-SVseries} shows the impact 
	of the constraining dataset on the principal components.
	Increasing the set of fit-observables by adding a new kind of data, when going from SV-E to SV-min and from SV-min to SV-bas, increases the kind of meaningful directions in the parameter space. The step from SV-min to SV-bas is particularly dramatic.
	Adding information
	on nuclear resonance properties to the dataset, increases the  number of
	relevant parameters to 6-7. The Skyrme functional is
	capable of describing  dynamical nuclear response; hence, its parameter space was not sufficiently probed  when tuning it to  ground state properties. Clearly, considering heterogeneous datasets is important for a balanced model optimization.
	Still, there is a significant room for improvement: the capabilities of
	the Skyrme functional are not yet fully explored by the extended dataset of SV-bas and more
	features are likely to  be accommodated. On the other hand, recent studies of isotopic
	shifts have demonstrated that the Skyrme functional is not
	flexible enough to describe the new kind of data
	\cite{Reinhard2017,Mil19a}. This calls
	for further model developments. Statistical analyses can be extremely helpful in such an undertaking as they elucidate the hidden features of a model.

	Comparing SV-min(NMP) with
	SV-min(t,x) one can see a dramatic effect from {\it the way} the functional
	is parametrized. Indeed, it is somehow astonishing that by replacing the traditional (t,x) form of Skyrme parameters  with a physically-motivated NMP input, reduces the span of eigenvalues by four orders of magnitude. Turning the
	argument around, we see that results of the principal component analysis
	depend sensitively on the way the model is
	formulated. If we were smart enough to guess all
	the ``physical'' parameter combinations, we could reduce the span of
	eigenvalues to less than one order of magnitude thus rendering each
	model parameter relevant. The step from the traditional Skyrme
	parametrization to the NMP-guided input was already such a physically motivated
	reduction. Still more may be possible.
	
	\section{Conclusions}
In this study, we applied a variety of statistical tools, both frequentist and Bayesian, to gain a deeper understanding of two commonly used nuclear mass models: the 4-parameter semi-empirical mass formula and the 14-parameter realistic Skyrme energy density functional.
In both cases, the principal component analysis shows that the effective number of degrees of freedom is  much lower. It is 1-2 for the LDM and 4-6 for the Skyrme functional. 

We studied the effect of the fitting domain on parameter estimation and correlation, and found it significant. While the values of optimal parameters may not change much in some cases, changing the fitting domain often results in a very different picture of correlations between parameters and/or observables. It is obvious, therefore, that
statements such as ``Quantity $A$ is strongly correlated with quantity $B$" must be taken with a grain of salt, as correlations not only depend on the model used but 
are also conditioned on the domain of fit-observables used to inform the model. In particular, using datasets containing strongly correlated homogeneous data (e.g., consisting of nuclear masses only) can result in spurious correlations and an incorrect physics picture.

We have seen that BMA can be advantageously employed to improve  predictions and uncertainty quantification for the LDM model, as observed in previous works for microscopic global mass models \cite{Neufcourt2019,Neufcourt2020,Neufcourt2020a}. 
Nevertheless an important limitation is the size of the domain on which ``reasonable'' evidence integrals can be obtained, otherwise BMA turns out to be a model selection. 
We recommend that evidences are evaluated on a reasonable number
of sampling datapoints (10 seems to be a practical upper bound when averaging 
state-of-the-art global nuclear mass models
\cite{Neufcourt2019,Neufcourt2020,Neufcourt2020a}).
BMA is also very sensitive to the nominal uncertainty of 
models, which needs to be tuned adequately to avoid 
numerical pitfalls.
When other methods to compute evidence integrals
become unrealistic,
the Laplace approximation remains a reliable and manageable alternative. 
Also, the Nested Sampling algorithm proposed in Ref.~\cite{NestedSampling} and expanded by Ref.~\cite{MultiNest}
offers another way to potentially address some of these issues.

Turning to the Skyrme model, we have noticed that
the principal component analysis and the effective number of degrees of freedom depend on the way the model is formulated. This speaks in favor of using parameters linked to physically-motivated quantities.

We believe that the use of rather standard statistical methodologies and diagnostic tools advocated in this work will be useful in further studies of nuclear models, both for the sake of understanding their structure and for practical applications.

	%%%%% Acknowledgements %%%%%%%
	
	\bigskip  
	Useful discussions with Earl Lawrence, Stefan Wild, and Samuel Giuliani are gratefully appreciated.
	This material is based upon work supported by the U.S.\ Department of Energy, Office of Science, Office of Nuclear Physics under award numbers DE-SC0013365 (Michigan State University) and DE-SC0018083 (NUCLEI SciDAC-4 collaboration).

	\section*{References}
%\bibliographystyle{naturemag}
%\bibliography{Bayes-masses}

\end{document}